\documentclass[12pt]{article}
\pdfoutput=1

\usepackage{color}
\usepackage{amsmath}
\usepackage{amsfonts}
\usepackage{amssymb}
\usepackage{graphicx}
\usepackage{subfig}             

\usepackage{mleftright}

\usepackage[english]{babel}
\usepackage[applemac]{inputenc}
\usepackage{braket}
\DeclareMathOperator{\sech}{sech}

\usepackage{cite}

\usepackage[	colorlinks=true,	citecolor=black,	linkcolor=black,	urlcolor=blue,hypertexnames=false]{hyperref}

\def\beqar{\begin{eqnarray}}
\def\eeqar{\end{eqnarray}}

\def\beq#1\eeq{\begin{align}#1\end{align}}

\def\bean{\begin{equation*}}
\def\eean{\end{equation*}}
\newcommand{\arXiv}[2]{\href{http://arxiv.org/pdf/#1}{{\tt #2/#1}}}
\newcommand{\arXivold}[1]{\href{http://arxiv.org/pdf/#1}{{\tt #1}}}

\numberwithin{equation}{section} 

\newcommand{\Tr}{\mathrm{Tr~}}

\newcommand{\RN}[1]{%
  \textup{\uppercase\expandafter{\romannumeral#1}}%
}

\newcommand{\eps}{\epsilon}

\newcommand{\bsp}{\begin{split}}
\newcommand{\esp}{\end{split}}

\newcommand{\mn}{\mu\nu}

\newcommand{\gcmic}{\,\mathrm{g\,cm^{-3}}}
\newcommand{\erg}{\,\mathrm{erg}}

\newcommand{\keV}{\,\mathrm{keV}}
\newcommand{\MeV}{\,\mathrm{MeV}}
\newcommand{\GeV}{\,\mathrm{GeV}}
\newcommand{\GeVinv}{\,\mathrm{GeV^{-1}}}
\newcommand{\TeV}{\,\mathrm{TeV}}

\newcommand{\s}{\sigma}

\newcommand{\order}[1]{O(#1)}
\newcommand{\nn}{\nonumber}

\definecolor{MyBlue}{rgb}{0.1,0.1,0.8}

\begin{document}
\begin{titlepage}

\begin{center}

	{	
		\LARGE \bf 
		The Light Radion Window
	}
	
\end{center}
	\vskip .3cm
	
	\renewcommand*{\thefootnote}{\fnsymbol{footnote}}

%

\begin{center} 
{\bf \  Fayez Abu-Ajamieh\footnote{\tt \scriptsize
		 \href{mailto:abuajamieh@ucdavis.edu}{abuajamieh@ucdavis.edu}
		 }
, Jun Seok Lee\footnote{\tt \scriptsize
		 \href{mailto:phylee@ucdavis.edu}{phylee@ucdavis.edu}
		 }
, and John Terning\footnote{\tt \scriptsize
		  \href{mailto:jterning@gmail.com}{jterning@gmail.com}
		 }
		} 
\end{center}

	\renewcommand{\thefootnote}{\arabic{footnote}}
	\setcounter{footnote}{0}


\begin{center} 

	{\it Department of Physics, University of California Davis\\One Shields Ave., Davis, CA 95616}

\end{center}


\centerline{\large\bf Abstract}

\begin{quote}
Inspired by the Contino-Pomarol-Rattazzi mechanism we explore scenarios with a very light (1 keV to 10 GeV) radion which could be associated with the suppression of the electroweak contribution to vacuum energy. We construct explicit, realistic models that realize this mechanism and explore the phenomenological constraints on this class of models. Compared with axion-like particles in this mass range, the bounds from SN 1987a and from cosmology can be much weaker, depending on the mass of the radion and its coupling to other particles.  With couplings suppressed by a scale lower than 100 TeV, much of the mass window from 100 keV to 10 GeV is still open.
\end{quote}

\end{titlepage}


\section{Introduction}

The hierarchy problem---the quantum instability of the weak scale ($\sim 10^{3}$ GeV) with respect to the Planck scale ($\sim 10^{19}$ GeV)---is a long-standing stumbling block in particle physics. One interesting class of models, based on the Randall-Sundrum (RS) model~\cite{Randall:1999ee},  uses a warped extra dimension in order to generate a stable hierarchy of scales. In these models, two 3-branes, the UV and IR branes, are embedded in  anti-de Sitter (AdS) space:
\beq
ds^{2} = e^{-2ky}\eta_{\mu\nu}dx^{\mu}dx^{\nu} - dy^{2}\, ,\label{RS metric}
\eeq
where the UV brane is located at $y=0$ and the IR brane is located at $y=\pi r_{c}$, where $r_{c}$ is the ``radius of compactification", $k$ is the inverse of the AdS curvature radius and an $S^{1}/Z_{2}$ symmetry is assumed so that both branes are stable. The electroweak scale (set by the location of the IR brane) is thus suppressed relative to  the Planck scale or UV scale through the exponential warping of the metric. The RS model provides a simple escape from the hierarchy problem, however, the original model contained two fine-tunings that relate the energy densities on the branes to the bulk cosmological constant. One fine-tuning is required to arrange for the correct separation of the two branes, while a second fine-tuning is needed to ensure a vanishing 4D cosmological constant. One could easily over-look the fine-tuning of the cosmological constant, since currently all models of particle physics make the same fine-tuning, but fine-tuning of the IR brane energy density is problematic for the following reason: it plays a direct role in determining the hierarchy between the electroweak and UV scales, and if the effective potential is really independent of the brane separation it means that the AdS space is unstable to fluctuations in the size of the extra dimension \cite{Csaki:1999jh}, corresponding to a massless particle, the radion. A massless radion  produces a long range force that couples to the trace of the stress-energy tensor.

To overcome this issue, Goldberger and Wise (GW)  \cite{Goldberger:1999uk} proposed a mechanism to stabilize the size of the extra dimension, thus generating a mass for the radion. In the GW mechanism, a bulk scalar sector is added, and the competition between the scalar's extra dimensional gradient and the conflicting boundary conditions produces an effective potential that stabilizes the size of the extra dimension. In the RS model, the radion plays the role of the Goldstone boson associated with  Spontaneous Breaking of Scale Invariance (SBSI), aka the dilaton.  The GW mechanism provides an explicit breaking of scale invariance and thus the radion generically becomes a Pseudo-Nambu-Goldstone Boson. The Goldstone nature of the radion/dilaton partially determines its coupling to standard model fields, resulting in interesting phenomenological signatures \cite{Csaki:1999mp,Chacko:2013dra,Bunk:2017fic,Foot:2007iy,Foot:2011et,Kobakhidze:2017eml,Arunasalam:2017ajm}.

In general, obtaining a light dilaton requires keeping any explicit breaking of scale invariance  small, so $\beta$ functions  associated with the approximately scale invariant sector must remain small over a range of scales. This is due to the fact that scale invariance allows for a non-derivative self-interaction quartic term for the dilaton, which can  actually prevent SBSI \cite{Bellazzini:2013fga}. In the context of the  RS model, a negative quartic effective potential would result in an unbounded negative energy and a runaway vacuum state,  while a positive quartic effective potential would result in a vanishing vacuum expectation value (VEV), so that scale invariance is not broken at all. These two disasters can be avoided if the quartic coupling has additional dependence on the radion, which can arise through slowly running couplings \cite{CPR,Bellazzini:2013fga,Coradeschi:2013gda,Cleary:2015pva,Megias:2015qqh,Agrawal:2016ubh}. This is the Contino-Pomarol-Rattazzi mechanism \cite{CPR}. In this scenario, a small quartic is present, but there is a non-trivial minimum due to a small amount of running. One can work out how this can lead to the potential being almost zero at its minimum.

More specifically, classically the effective potential of a dilaton is \cite{Fubini:1976jm}
\beq \label{quartic}
V_{eff} = \Lambda \chi^{4}\, ,
\eeq
where $\chi$ is a dimensionless field which parameterizes a non-linear realization of the dilaton, $\sigma$ by:
\beq
\chi=e^{\sigma/f}~.
\eeq
Under a scale transformation, and operator ${\mathcal O}$ of dimension $\Delta$ transforms by
\beq
{\mathcal O}(x) \to \rho^\Delta {\mathcal O}(\rho x)~,
\eeq
while
\beq
\sigma(x)\to \sigma(\rho x)+f \ln \rho~.
\eeq

Classically, in order for SBSI to occur, one needs to tune $\Lambda$, a contribution to the vacuum energy, to zero. However, if a slowly-running perturbing operator is introduced, then the running can lead to a dependence of $\Lambda$ on $\chi$, and a suppressed value of $V_{eff}$ at a non-trivial minimum that corresponds to SBSI:
\beq\label{F_vanishing}
\frac{d \Lambda(\lambda(\chi)) }{d\chi} \chi+ 4 \Lambda(\lambda(\chi))  = 0.
\eeq
As shown in \cite{Bellazzini:2013fga,Coradeschi:2013gda},  this can be achieved by the introduction of an almost marginal operator with dimension $4-\epsilon$ that explicitly breaks scale invariance. The running of the coupling, $\lambda$, of this operator satisfies
\beq
\beta(\mu) = \epsilon \, b(\lambda(\mu)) \ll 1
\eeq
which means that the first term in Eq. (\ref{F_vanishing}) is of order $\epsilon$, so at the minimum, with $\sigma=0$, we also have $V_{eff}$  of order $\epsilon$. This gives a mass squared for the dilaton of order  $\epsilon f^2\sim \epsilon \Lambda^{1/2}$. 

Explicit 5D  \cite{Bellazzini:2013fga,Coradeschi:2013gda} and 4D models \cite{Cleary:2015pva} that incorporate the Contino-Pomarol-Rattazzi mechanism have been constructed. In this paper, we investigate this class of models focussing on the case when the radion mass is between 100 keV and 10 GeV. Somewhat surprisingly, there is an open window that is not currently ruled out. These models have the additional interesting property that if the Electroweak sector of the standard model is part of the approximately conformal sector that gives rise to the dilaton, as happens in RS models, then the Electroweak contribution to the vacuum energy can be suppressed by orders of magnitude; it can even be of the order the QCD contribution to the vacuum energy \cite{Bellazzini:2013fga}.

The goals of this paper are to discuss a range of realistic models and to examine the phenomenological constraints. It will prove helpful to frame the discussion in terms of how the well-known bounds on Axion-Like-Particles (ALPs) are modified in the case of light radions.

A brief outline of the paper is as follows.
We construct a realistic 5D model  in Section 2 and show that it can predict a radion mass far below 10 GeV in Section 3. We calculate the coupling of the radion to standard model particles in Section 4. Special attention is given to the coupling to massless gauge bosons, and we review how this coupling is actually model-dependent~\cite{Bellazzini:2012vz}. This model dependence can drastically modify the light radion search limits. The couplings to photons and gluons are important since they can lead to large nucleon couplings \cite{Burgess:2000yq,Andreas:2008xy,Kribs:2009fy,Cheng:2012qr,Cheng:2014}.  Readers only interested in the phenomenological aspects can skip directly to section \ref{Chap:Limits} where we discuss the experimental constraints on very light radions, particularly from astrophysical observations. We close out by presenting some brief conclusions and give a summary plot of the open window in section \ref{Chap:Conclusions}.
 We also provide a brief review of 5D theories with bulk gauge bosons in Appendix~\ref{massivegbmass}, and provide examples of benchmark 5D  models with parameter values consistent with a very light radion in Appendix~\ref{Chap:Results}.

\section{A Light Radion via a Small $\beta$-function}\label{Chap:IRmistuning}
\setcounter{equation}{0}

We begin by quickly reviewing the model of ref. \cite{Bellazzini:2013fga}. The 5D action is given by:
\beq\label{action}
S =\int dx^{5} \sqrt{g} \Big(-\frac{1}{2\kappa^{2}} \mathcal{R} +\frac{1}{2}g^{MN}\partial_{M}\phi \partial_{N}\phi \hspace{1 mm} - \hspace{1 mm} V(\phi)  \Big)\ -\sum_{i=0,1} \int dx^{4} \sqrt{g_{i}}V_{i}(\phi).
\eeq
where $\phi$ is a bulk scalar, $\kappa^{2}$ is the 5D Newton constant and $g_{0,1}$ are the induced metrics on the UV and IR branes respectively. The brane localized potentials $V_{0,1}$ are chosen to be:
\beq \label{braneV}
V_{i}({\phi}) = T_{i} + \lambda_{i}(\phi-v_{i})^{2}.
\eeq
A 4D Lorentz invariant solution to the Einstein equations can be found, and we take the metric to be:	
\beq\label{bgmetric}
ds^2=e^{-2A(y)}\eta_{\mu\nu}dx^{\mu}dx^{\nu}-dy^2\, ,
\eeq
where $e^{-A(y)}$ is the general warp factor and the UV (IR) brane is placed at $y_{0} (y_{1})$.  Greek indices only run over ordinary 4 dimensional spacetime throughout this paper. The solution of the Einstein equations,
\beq
R_{ab} = \kappa^2 \tilde{T}_{ab}=\kappa^2\left(T_{ab}-\frac{1}{3}g_{ab}g^{cd}T_{cd}\right)~,
\eeq
gives the  following equations of motion for the warp factor and scalar field:
\beq \label{EOM1}
4{A'}^{2} - A^{\prime \prime}  &= -\frac{2\kappa^{2}}{3}V(\phi)\, ,\\
{A'}^{2} &= \frac{\kappa^{2}{\phi'}^{2}}{12} - \frac{\kappa^{2}}{6}V(\phi)\, ,\\
\phi^{\prime \prime}  &= 4A'\phi'+ \frac{\partial V}{\partial \phi}\, ,
\eeq
and boundary conditions:
\beq 
2A'|_{y_{0,1}} =& \pm \frac{\kappa^{2}}{3} V_{1}(\phi)|_{y_{0,1}}\, ,\label{BC1}\\
2\phi'|_{y_{0,1}} =& \pm \frac{\partial V_{1}}{\partial \phi} |_{y_{0,1}}\, ,\label{BC2}
\eeq
where the $+$ sign is for the UV brane and the $-$ sign is for the IR brane.

The bulk scalar potential includes a constant term that represents the bulk cosmological constant, and a mass term, which parametrizes the small renormalization group running $\epsilon$ of the 4D CFT. Thus the potential is given by:
\beq \label{bulkV}
V(\phi) = - \frac{6k^{2}}{\kappa^{2}} -2\epsilon k^{2}\phi^{2}~,
\eeq
where $k$ is the asymptotic AdS curvature scale. The approximate solution of the equations of motion is:
\beq
A(y) &= -\frac{1}{4} \log \Big[ \frac{\sinh(4k(y-y_{c}))}{\sinh(-4ky_{c})}  \Big] \label{bgsolmetric}\, ,\\
\phi(y) &= v_{0} e^{\epsilon k(y-y_{0})} -\frac{\sqrt{3}}{2} \log (\tanh(2k(y_{c}-y)))\, , \label{bgsolphi}
\eeq
where  $y_{c}$ parameterizes the scale of the condensate developed by the running of the perturbing operator. This condensate is shielded by the IR brane, so $y_{c} > y_{1}$. The effective dilaton potential can be found by using the solutions to integrate out the bulk scalar in favor of its boundary values. The effective potential receives contributions from both boundaries:
\beq \label{effectiveUV/IR}
V_{UV/IR} = e^{-4A(y_{0,1})} \Big[V_{0,1}(\phi(y_{0,1}))\mp \frac{6}{\kappa^{2}}A'(y_{0,1})  \Big].
\eeq
If we rewrite (\ref{effectiveUV/IR}) in term of the dilaton field \cite{Bellazzini:2013fga} 
\beq
\chi= e^{-A(y_1)}~, 
\eeq
we find that the IR effective potential has the form:
\beq\label{effectiveDilatonV}
V_{IR} = \chi^{4} \Big[  V_{1} \big(\phi(A^{-1} (-\log\chi)) \big) + \frac{6}{\kappa^{2}} A'\big(A^{-1} (-\log\chi) \big) \Big]
\eeq
which has the required form of (\ref{quartic}). Comparing (\ref{effectiveDilatonV}) with (\ref{quartic}), we immediately identify the coefficient of the radion quartic term (i.e. the contribution to vacuum energy) as:
\beq\label{F}
\Lambda = V_{1} +\frac{6}{\kappa^{2}} A'.
\eeq
The effective potential is obtained by substituting the bulk solutions into the bulk action and integrating over the extra dimension. In the $\lambda_{0,1}\to\infty$ limit, this yields two boundary terms:
\beq \label{UVeffectiveV}
V_{UV} = \mu_{0}^{4} \Big[ T_{0} - \frac{6k}{\kappa^{2}} \Big]\, ,
\eeq
\beq \label{VIR}
V_{IR} = \chi^{4} \Big[ T_{1} + \frac{6 k}{\kappa^{2}} \cosh \Big( \frac{2 \kappa}{\sqrt{3}} (v_{1} - v_{0}(\mu_{0}/\chi)^{\epsilon}) \Big) \Big] \sech^{2} \Big( \frac{\kappa}{\sqrt{3}}(v_{1} - v_{0}(\mu_{0}/\chi)^{\epsilon}) \Big)\, ,
\eeq
where $\mu_{0} = e^{-k y_{0}}$ and $\chi =  e^{-k y_{1}}$ parameterize the locations of the UV and IR branes respectively.  The UV brane potential is just a constant that is tuned to zero, this is just the usual UV RS tuning. On the other hand, the IR potential has a nontrivial minimum which determines the size of the extra dimension, and the scale of SBSI. The vacuum of IR potential is obtained from \eqref{VIR}. It reads
\beq\label{VIRmin}
V^{min}_{IR}=-\epsilon\frac{6\sqrt3 k v_0}{\kappa}\tanh(\frac{\kappa}{\sqrt{3}}(v_1-v_0(\mu_0/\chi)^{\epsilon})\left<\chi\right>^{4}(\mu_0/\chi)^{\epsilon}),
\eeq
where we can explicitly see the suppression factor $\epsilon$.



\section{The Radion Mass}\label{Chap:MassandVacuum}
\setcounter{equation}{0}

In order to canonically normalize the dilaton, we need to properly include the metric fluctuations that mix with the Goldberger-Wise field $\phi$.
For a general ansatz to describe the fluctuations, we will follow the derivation and conventions in refs. \cite{Csaki:2000zn} and \cite{Csaki:2007ns}. The fluctuating metric is 
\beq\label{perturbedmetric}
ds^2=e^{-2A(y)-2F(x,y)}\eta_{\mu\nu}(x)dx^{\mu}dx^{\nu}-(1+G(x,y))^2dy^2\, ,
\eeq
where $F(x,y)$ and $G(x,y)$ are the small fluctuations. We decompose the scalar into a background profile and fluctuations as
\beq
\phi(x,y)=\phi_0(y) + \varphi(x,y),
\eeq
where $\phi_0$ is the bulk solution (\ref{bgsolphi}).
The linearized Einstein equations are
\beq
\delta R_{ab} = \kappa^2 \delta \tilde{T}_{ab}.
\eeq
First, the linearized equation for $\delta R_{\mn}$ gives $G(x,y)=2F(x,y)$ (see ref. \cite{Csaki:2000zn}). Then the linearized Einstein equation for each $\delta R_{\mn}$, $\delta R_{\mu5}$, and $\delta R_{55}$ are
\beq
\delta R_{\mn}&=\eta_{\mu}\Box F+e^{-2A}\eta_{\mn}(-F^{\prime \prime} +10A^\prime F^\prime+6A^{\prime \prime} F-24A^{\prime 2}F), \\
\delta R_{\mu5}&=3\partial_{\mu}F^\prime-6A^\prime\partial_{\mu}F, \\
\delta R_{55}&=2e^{2A}\Box F+4F^{\prime \prime} -16A^\prime F^\prime 
\eeq
and the source terms are
\beq
\delta \tilde{T}_{\mn}&=-\frac{2}{3}e^{-2A}\eta_{\mn}(V^\prime (\phi_0)\varphi-2V(\phi)F) \nn\\
&-\frac{1}{3}e^{-2A}\eta_{\mu}\sum\limits_i\left(V^\prime_i(\phi_0)\varphi-4V_i(\phi)F\right)\delta(y-y_i),\\
\delta \tilde{T}_{\mu5}&=\phi_0^\prime \partial_{\mu}\varphi, \\
\delta \tilde{T}_{55}&=,2\phi_0^\prime \varphi^\prime +\frac{2}{3}V^\prime (\phi_0)\varphi+\frac{8}{3}V(\phi_0)F  \nn \\ 
&+\frac{4}{3}\sum\limits_i\left(V^\prime_i(\phi_0)\varphi+2 V_i(\phi_0)F\right)\delta(y-y_i)
\eeq
The linearized equation of motion for the scalar field is
\beq
e^{2A} \Box \varphi - \varphi ^{\prime \prime}  + 4 A^\prime  \varphi ^\prime  + 
V^{\prime\prime}(\phi_0) \varphi
=&-6 \phi_0 ^\prime  F^\prime  -4 \frac{\partial V}{\partial \phi} F \nn \\
& 
-\sum_i \left(V_i^{\prime\prime}(\phi_0)  \varphi
+2  V^\prime_i(\phi_0) F \right)\delta(y-y_i).
\eeq
The Einstein equation for $\delta R_{\mu5}$ can be immediately integrated to give the coupled equation
\beq\label{5mu}
\phi_0^\prime\varphi=\frac{3}{\kappa^2}(F^\prime -2A^\prime F)~.
\eeq
The boundary equation from Einstein equation that is non-redundant is \cite{Csaki:2000zn}
\beq\label{phiprimebd}
[\varphi ^\prime ]|_i=V^{\prime \prime}_i(\phi_0)\varphi+V^\prime_i(\phi_0)F~.
\eeq
where $i=0,1$ again corresponds to UV, IR branes respectively.
Considering the combination
\beq
\frac{1}{4}e^{2A}\eta^{\mn}\delta R_{\mn}+\delta R_{55}
\eeq
gives
\beq
e^{2A}\Box F+F^{\prime \prime} -2A^\prime F^\prime =\frac{2\kappa^2}{3}\phi^\prime _0\varphi^\prime ,
\eeq
and using the equation \eqref{5mu}, we get the bulk equation which only involves the fluctuation $F$ and the background solution of the metric and the scalar field,~\eqref{bgsolmetric}-\eqref{bgsolphi}:
\beq\label{bulkmasseq}
F^{\prime \prime} -2A^\prime F^\prime -4A^{\prime \prime} F-2\frac{\phi^{\prime \prime}_0 }{\phi^\prime_0 }F^\prime +4A^\prime \frac{\phi^{\prime \prime}_0 }{\phi^\prime_0 }F=e^{2A}\Box F~.
\eeq
Together with the boundary condition \eqref{phiprimebd}, we can use this equation to determine the Kaluza-Klein (KK) eigenmodes and mass eigenvalues for $F$, since the eigenmodes satisfy
\beq
\Box F = -m^2 F~.
\eeq
The equation for the mass eigenvalue of the equation \eqref{bulkmasseq} can be solved numerically. In \cite{Bellazzini:2013fga}, the mass squared of the radion  was found to be linear in $\epsilon$, to the leading order in $\epsilon$, which our numerical solutions confirm. As discussed later in Sec.~\ref{Chap:Limits} we are interested mostly in the 100 keV to 10 GeV mass range of radion which corresponds to $\epsilon$ in the range of 	$10^{-17}$ to $10^{-11}$.  Examples of benchmark parameter values that yield such a light radion are given in Appendix~\ref{Chap:Results}.

Thus the mass of the radion/dilaton can be made small as long as the explicit scale invariance breaking $\epsilon$, is kept small, which corresponds to a very slow running of the coupling. In addition, the value of the IR potential at the minimum, which represents a contribution to vacuum energy, is also suppressed by $\epsilon$, so the Electroweak vacuum energy can be significantly reduced, even to be roughly the same size as the QCD contribution. 

The desired hierarchy and the effective potential minimum are obtained by controlling $v_0$ and $v_1$, which are the UV and IR values of the scalar field in the $\lambda_{0,1}\to\infty$ limit. We give the detailed results in Appendix \ref{Chap:Results}. Typically the ratio, $v_0/v_1$, is $O(10^{-1})$ for all the parameter range we study.

We also note that the coupling to SM fermions can give rise to radiative corrections to the radion mass.  We can estimate this correction through Naive Dimensional Analysis (NDA) to be approximately $\delta m^{2} \sim \frac{1}{16 \pi^{2}} m_f^2 g_{\sigma f f}^{2} \Lambda^{2}$. With cutoff scale $\Lambda\sim$ TeV, this corresponds $m_f g_{\s ff}<10^{-2}(m_\s/\mathrm{GeV})$ for the radiative mass correction to be negligible. For example, the $(g-2)_e$ constraints which we will discuss in Sec.~\ref{Chap:LimitsandDecays} gives $g_{\s ee}<10^{-2}\GeVinv$ for $m_\s=1\MeV$. Then we have  $10^{-2}(m_\s/\mathrm{GeV})\sim 10^{-5}$ whereas $m_e g_{\s ee}<5\times10^{-6}$, so the radiative corrections can be small.

Contino, Pomarol, and Rattazzi originally suggested \cite{CPR} that $\phi$ could be a 5D Goldstone boson and that $\epsilon$ could be an arbitrary parameter that breaks the corresponding symmetry. However, without a complete model in hand, the low-energy theory certainly seems to be fine-tuned. We will nevertheless proceed to examine the phenomenology of this model in spite of the fine-tuning issues, as one does for the standard model.

\section{Radion Couplings to Matter}\label{Chap:cp}


\subsection{Coupling to Brane Localized Fields}\label{branecoupling}

For the metric~\eqref{perturbedmetric} with the solution $F(x,y)=2G(x,y)$,
the perturbed term at linear order in $F$ is \cite{Csaki:2000zn,Csaki:2007ns}
\beq\label{delds}
\delta(ds^2)=-2F(e^{2A}\eta_{\mu\nu}dx^{\mu}dx^{\nu}+2dy^2).
\eeq
Then the linear term in the action is
\beq
S_{radion}=&-\frac{1}{2}\int d^5x\sqrt{g}\left(-2\frac{1}{\sqrt g}\frac{\delta {\mathcal L}}{\delta g^{MN}}\right)\delta g_{MN}\\
=&-\frac{1}{2}\int d^5x\sqrt{g}T^{MN}\delta g_{MN}, \label{Tdelg}~
\eeq
where $\delta g_{MN}$ is given by \eqref{delds} and for fields localized on the UV/IR brane,
\begin{equation*}
\begin{split}
S_{radion} \supset \int d^4x\sqrt{g_1}F(x,y_{0,1})\mathrm{Tr}T_{\mu\nu}~,
\end{split}
\end{equation*}
where $y=y_{0,1}$ corresponds to the UV/IR brane respectively. Thus, we get a tree level radion coupling to fields on the brane:
\beq\label{braneLambda}
\tilde{F}(y_{0,1})\sigma(x)\Tr T_{\mu\nu}\equiv \frac{1}{\Lambda_{UV/IR}}\sigma(x)\Tr T_{\mu\nu}~,
\eeq
where we have factored the fluctuation as $F(x,y)=\tilde{F}(y)\sigma(x)$, where $\tilde{F}(y)$ is the lightest KK eigenmode from (\ref{bulkmasseq}) and  $\sigma(x)$ is a canonically normalized 4D radion field. The fluctuations $F(x,y)$ and $\varphi(x,y)$ are related by Equation~\eqref{5mu},
which implies the decomposition of $\varphi$ with the same 4D radion field $\sigma(x)$,
\begin{equation*}
\varphi(x,y)=\tilde{\varphi}(y)\sigma(x),
\end{equation*}
where
\beq\label{varphieq}
\tilde{\varphi}=\frac{3}{\kappa^2}\frac{(\tilde{F}^\prime -2A^\prime \tilde{F})}{\phi_0^\prime }.
\eeq

While the solution $F(x,y)$ is obtained from \eqref{bulkmasseq}, its overall normalization depends on the canonical normalization of the radion kinetic term which has two contributions, namely from the metric fluctuation and from the bulk scalar field. Expanding the Ricci scalar up to the second order in $F$,
\beq
-\frac{1}{\kappa^{2}} \int_{y_{0}}^{y_{1}} dy \sqrt{g} \mathcal{R}=\frac{1}{\kappa^2}\int^{y_1}_{y_0}dy ~e^{-2A(y)}\left(6(\partial F)^2+O(F^3)\right)~,
\eeq
where the orbifolding factor of 2 is included, so the gravity contribution is
\beq
\mathcal{L}^{(kin)}_{eff}&\supset\frac{1}{\kappa^2}\int^{y_1}_{y_0}dy ~e^{-2A(y)}\left(6(\partial F)^2\right)~.
\eeq
From the bulk scalar kinetic term we find
\beq
\int^{y_1}_{y_0}dy~ e^{-2A(y)}(\partial (\phi+\varphi))^2~,
\eeq
and the orbifolding factor of 2 is also included, so the bulk scalar contribution is
\beq
\mathcal{L}^{(kin)}_{eff}&\supset \int^{y_1}_{y_0}dy~ e^{-2A(y)}(\partial \varphi)^2.
\eeq

To canonically normalize the radion field $\sigma(x)$, the bulk wave functions $\tilde{F}$ and $\tilde{\varphi}$ should satisfy
\beq\label{normal}
\int^{y_1}_{y_0}dy \left( e^{-2A(y)}\frac{6}{\kappa^2}(\tilde{F}(y))^2 + e^{-2A(y)}(\tilde{\varphi} (y))^2 \right)=\frac{1}{2}.
\eeq
With this normalization the  action is:
\beq
\mathcal{S}^{(eff)}_{radion}=\int d^4x\left(\frac{1}{2}\, \partial^\mu \sigma \partial_\mu \sigma -V_{eff}(\sigma)+\sqrt{g_1} \frac{\sigma(x)}{\Lambda_{IR}}\,T_{(IR) \mu}^{\mu}+\sqrt{g_0}\frac{\sigma(x)}{\Lambda_{UV}}\,T_{(UV) \mu}^{\mu}\right)~.
\eeq
As expected the coupling of the radion to fields on the UV brane is suppressed by $\Lambda_{UV}$ while the coupling to the IR brane is suppressed by 
$\Lambda_{IR}$.

\subsection{Coupling to Massless Gauge Bosons}\label{bulkcoupling}

The coupling of the radion to massless gauge bosons is loop-induced and is quite model-dependent. The radion coupling to gauge fields in the bulk includes, in addition to the overlap between the wavefunctions of the radion and gauge boson, a contribution from the  trace anomaly. 
To see how the radion couples to the massless bulk gauge fields \cite{Bellazzini:2012vz,Csaki:2007ns}, it is simplest to look at the full matching of the gauge coupling, renormalized at a scale $\mu$:
\beq\label{gaugecouplingrun}
\frac{1}{g^2(\mu)}=\frac{ R \log(\frac{\mu_0}{f})}{g_5^2}- \frac{b_{IR}}{8\pi^2}\log\left(\frac{f}{\mu}\right)-
 \frac{b_{elem}}{8\pi^2}\log\left(\frac{\mu_0}{\mu}\right)\, ,
\eeq
where $R=2/k$ is the AdS curvature with the orbifolding included, while $\mu_0$ and $f$ represent the energy scales of the UV and IR branes. The first term comes from the bulk tree-level contribution which corresponds to the CFT contribution to the running  (see Appendix~\ref{massivegbmass}), so  we can identify
\beq
b_{CFT} = -\frac{ 8 \pi^2 R }{g_5^2}.
\eeq
The second term in (\ref{gaugecouplingrun}),  $b_{IR}$, is the $\beta$ function coefficient due to  IR localized fields which are lighter than $\mu$. The third term, $b_{elem}$, is the $\beta$ function coefficient due to UV localized fields which correspond to elementary fields weakly coupled  to the CFT\footnote{In \cite{Csaki:2007ns} this contribution is denoted as $b_{UV}$.}. 

We can find the radion coupling by looking at the effective gauge action:
\beq
{\mathcal L}_{AA} =-\frac{1}{4 \,g^2(\mu)} G^a_{\mu\nu}G^{a\mu\nu} ~.
\eeq
The radion field can be thought of as the fluctuation of the IR brane, therefore the radion coupling to the gauge field can be obtained \cite{Bellazzini:2012vz,Csaki:2007ns} by substituting $f \rightarrow f e^{\sigma/f}$. So we find the coupling:
\beq\label{cptogluon}
{\mathcal L}_{\sigma AA}= \frac{g^2}{32\pi^2} \left(b_{IR}-b_{CFT} \right) \frac{\sigma}{f} \,G^a_{\mu\nu}G^{a\mu\nu} \, ,
\eeq
where we have returned to canonically normalized gauge fields.

Thus the coupling of the radion to a gauge field is a completely model-dependent parameter. For example, consider the coupling to the gluon; the two $\beta$ function coefficients in (\ref{cptogluon}) depend on which colored fields are composites of the approximate conformal sector.  An important special case is when all of the colored fields are elementary, i.e. localized on the UV brane. In this case there is no direct coupling of the radion to gluons.


\subsection{Coupling to Nucleons through Gluons}\label{subsec:nucleoncoupling}
The contribution to the effective coupling to nucleons comes from quarks and gluons. This calculation has been done for the Higgs \cite{Burgess:2000yq,Andreas:2008xy,Kribs:2009fy,Cheng:2012qr,Cheng:2014}, and we can follow a similar argument. 

The gluon and quark mass terms in the trace of the 4D energy momentum tensor are
\beq\label{EMtensor}
\Theta^{\mu}_{\mu}=m_u\bar{u}u+m_d\bar{d}d+m_s\bar{s}s+\sum\limits_{Q=c,b,t} m_Q\bar{Q}Q+\frac{\beta(g)}{2g}G^aG^a+...,
\eeq
The low-energy $\beta$-function of the gauge field can be obtained directly from \eqref{gaugecouplingrun}:
\beq\label{QCDbeta}
\frac{\beta(g)}{2g}G^a_{\mu\nu}G^{a\mu\nu}= \frac{1}{2g}\frac{\partial g}{\partial \log \mu}GG=-\frac{(b^{(3)}_{elem}+b^{(3)}_{IR})}{32\pi^2}g^2 G^a_{\mu\nu}G^{a\mu\nu}.
\eeq
The heavy quark expansion~\cite{Shifman:1978zn},
\beq\label{hqexpansion}
\sum\limits_{Q=c,b,t} m_Q\bar{Q}Q \rightarrow 3\times\left(-\frac{2}{3} \frac{g^2}{32 \pi^2}G^a_{\mu\nu}G^{a\mu\nu}\right)+O\left(\frac{1}{m_Q^2} \right),
\eeq
means that at leading order the stress tensor is independent of the $c$, $b$, $t$ quark terms, so 
\beq \label{stresstensor}
\Theta^{\mu}_{\mu}=m_u\bar{u}u+m_d\bar{d}d+m_s\bar{s}s-\frac{b^{(3)}_{light}}{32\pi^2}g^2 G^a_{\mu\nu}G^{a\mu\nu}+...,
\eeq
where the $\beta$ function coefficient $b^{(3)}_{light}$ includes only the $u$, $d$,  $s$ quarks and the gluon which we assume are all elementary. From \eqref{QCDbeta} and \eqref{hqexpansion} it is
\beq
b^{(3)}_{light}=(b^{(3)}_{elem}+b^{(3)}_{IR})+2\,.
\eeq
The nucleon mass is effectively given by the matrix element of the trace of the energy momentum tensor at vanishing momentum transfer,
\beq\label{nucleonmass}
m_N\bar{N}N=\Braket{N|\Theta^{\mu}_{\mu}|N}.
\eeq
The radion couples to the nucleons through the gluon coupling \eqref{cptogluon}, and neglecting the contributions from the light quarks' masses we find the radion-nucleon coupling to be
\beq\label{nucleoncoupling}
g_{\sigma NN}m_N \sigma \bar{N} N \equiv \Braket{N| \frac{g^2}{32 \pi^2} \left(b_{IR}^{(3)} -b^{(3)}_{CFT}\right)  \frac{\sigma}{f} \, G^a_{\mu\nu}G^{a\mu\nu} |N}  
=\frac{b^{(3)}_{CFT} -b^{(3)}_{IR}  }{b^{(3)}_{light}}\frac{m_N}{f} \sigma \bar{N}N\,.
\eeq

Notice that if the gluon and quarks are elementary, i.e. localized on the UV brane, then this leading contribution vanishes, and the radion coupling is suppressed by the scale of the UV brane (as seen from (\ref{braneLambda}) rather than by $f$).
When the radion coupling to both quarks and gluons is negligible, the radion can still couple to nucleons through photons, i.e. through the photon term in the stress tensor:
\beq
\Theta^{\mu}_{\mu}\supset -\frac{b^{EM}_{elem}+b_{IR}^{EM}}{32\pi^2}e^2 F_{\mu\nu}F^{\mu\nu}\, ,
\eeq
where $e$ represents the electromagnetic gauge coupling. Lattice calculations provide the best estimate of the QED contribution to the nucleon mass and up to NNNLO. The QED correction to neutron mass is calculated to be~\cite{Davoudi:2014qua},
 \beq
\frac{(\delta m_N)_{QED}}{m_N}\simeq 10^{-5}.
\eeq
Then we deduce
\beq\label{QEDcorrection}
\Braket{N|-\frac{b^{EM}_{elem}+b_{IR}^{EM}}{32\pi^2}e^2 F_{\mu\nu}F^{\mu\nu}|N}\simeq 10^{-5}\Braket{N|-\frac{b^{(3)}_{light}}{32\pi^2}g^2 G^a_{\mu\nu}G^{a\mu\nu}|N},
\eeq
and radion coupling to neutrons through the photon coupling, \eqref{cptogluon} is 
\beq\label{nucleoncouplingphoton}
g_{\sigma NN}m_N\sigma \bar{N} N &= \Braket{N| \frac{e^2}{32 \pi^2} \left(b_{IR}^{EM} -b^{EM}_{CFT}\right)  \frac{\sigma}{f} \, F^a_{\mu\nu}F^{a\mu\nu} |N}  \\
&\simeq \frac{\left(b^{EM}_{CFT}-b_{IR}^{EM} \right)}{\left(b^{EM}_{elem}+b_{IR}^{EM} \right)}10^{-5}\Braket{N| \frac{g^2}{32 \pi^2} \left(-b^{(3)}_{light}\right)  \frac{\sigma}{f} \, G^a_{\mu\nu}G^{a\mu\nu} |N} \\
&\simeq\frac{\left(b^{EM}_{CFT}-b_{IR}^{EM} \right)}{\left(b^{EM}_{elem}+b_{IR}^{EM} \right)}10^{-5}\frac{m_N}{f} \sigma \bar{N}N,
\eeq
where we used \eqref{QEDcorrection} and \eqref{nucleonmass} in the second and third lines. Writing the photon coupling term from \eqref{cptogluon} as
\beq\label{cptophoton}
{\mathcal L}_{\sigma AA}=&-\frac{1}{4}g_{\s\gamma\gamma} \,\sigma \,F_{\mu\nu}F^{\mu\nu},
\eeq
the correlation between coupling to photons and the coupling to nucleona reads
\beq
g_{\s \gamma\gamma}&\simeq10^{5}\,\frac{e^2(b^{EM}_{elem}+b_{IR}^{EM})}{8 \pi^2}\,g_{\s NN}\\
&\simeq1.16\times10^{2}\,(b^{EM}_{elem}+b_{IR}^{EM})\,g_{\s NN} .\label{gNggammacor}
\eeq
The phenomenology of this scenario is discussed further in subsection \ref{SN}.

\subsection{Radion Decay to Massive Particles}\label{radiondecay}

Before we discuss the experimental bounds on the radion's parameter space, we need to investigate the possibility of its decay into lighter particles, since this can also affect these bounds. In this subsection, we focus primarily  on the radion decay to massive particles, since the decay to photons has been extensively studied for the case of ALPs. If the radion decays quickly enough, then some of the experimental constraints are invalidated. For example if the radion decays in less than 1 second, the beginning of Big Bang Nucleosynthesis (BBN), then the constraints from cosmology will be lifted \cite{Kawasaki:2017bqm}.

For the mass range of interest ($m_{\sigma} \lesssim 10$ GeV), the radion can decay to fermions or mesons if the decay is kinematically allowed. The radion decay to two fermions is given by:
\beq\label{fermiondecay}
\Gamma(\sigma \rightarrow f\bar{f}) = \frac{1}{8\pi} m_{\sigma}m_f^2 g_{\sigma ff}^{2} \Big[ 1- \frac{4 m_{f}^{2}}{m_{\sigma}^{2}} \Big]^{3/2}~,
\eeq
where  $g_{\sigma ff}$ is the radion's low-energy, effective coupling to fermions:
\beq
{\mathcal L}_{\sigma ff} = g_{\sigma ff} m_f \s {\overline f} f~.
\label{effsigmaff}
\eeq
Thus $g_{\sigma ff}$ has units of inverse mass. 

The radion's coupling to mesons through quarks and gluons is similar to the case of nucleons. Focusing on decays to two pions, and denoting the invariant mass squared of two pions by $q^2$, the coupling to pions can be calculated as follows \cite{Voloshin:1985tc,Barbieri:1988ct} :
\beq
\Braket{\pi^+\pi^-|\Theta^{\mu}_{\mu}|0}&=q^2+2m_{\pi}^2~,\\
\Braket{\pi^+\pi^-|m_u\bar{u}u+m_d\bar{d}d+m_s\bar{s}s|0}&=m_{\pi}^2~,
\eeq
which implies
\beq
\Braket{\pi^+\pi^-|-\frac{b^{(3)}_{light}}{32\pi^2}g^2G^2|0}=q^2+m_{\pi}^2~.
\eeq
 Using \eqref{cptogluon} one obtains $\sigma\to\pi\pi$ decay amplitude,
\beq
A(\sigma\to\pi\pi)&= \frac{\left(b_{IR}^{(3)}-b_{CFT}^{(3)} \right)}{f}\Braket{\pi^+\pi^-|\frac{g^2}{32\pi^2} \,G^a_{\mu\nu}G^{a\mu\nu}|0},\\
&=-\frac{\left(b^{(3)}_{IR}-b^{(3)}_{CFT} \right)}{b^{(3)}_{light}}\frac{m_{\sigma}^2+m_{\pi}^2}{f}.
\eeq
Using \eqref{nucleoncoupling} one obtains the decay width,
\beq
\Gamma(\sigma\to\pi\pi)&=\frac{1}{8\pi m_{\sigma}^2}\left(\frac{m_{\sigma}^2}{4}-m_{\pi}^2\right)^\frac{1}{2} |A|^2 \\
&=\frac{g_{\sigma NN}^2}{16\pi} m_{\sigma}^3\left(1-\frac{4m_{\pi}^2}{m_{\sigma}^2}\right)^{\frac{1}{2}}\left(1+\frac{m_{\pi}^2}{m_{\sigma}^2}\right)^2~.\label{modelpiondecay}
\eeq
where $g_{\sigma NN}$ is given by \eqref{nucleoncoupling}. 
Using low-energy effective theory with a coupling
\beq\label{cptopion}
{\mathcal L}_{\sigma \pi\pi}=&g_{\s\pi\pi}\, m_\pi^2 \,\sigma \pi\pi,
\eeq
where $g_{\sigma \pi\pi}$ has units of inverse mass, one gets a decay width,
\beq\label{piondecay}
\Gamma(\sigma\to\pi\pi)=&\frac{g_{\sigma \pi\pi}^2m_\pi^4}{16\pi \,m_\sigma} \left(1-\frac{4m_{\pi}^2}{m_{\sigma}^2}\right)^{\frac{1}{2}}~.
\eeq
Comparing to \eqref{modelpiondecay} we find that
\beq\label{gpipigNN}
g_{\s\pi\pi}=g_{\s NN}\frac{m_\s^2 +m_\pi^2}{m_\pi^2}.
\eeq


\section{Limits}\label{Chap:Limits}
\begin{figure}[t]
 \centering
  \includegraphics[height=.45\textheight]{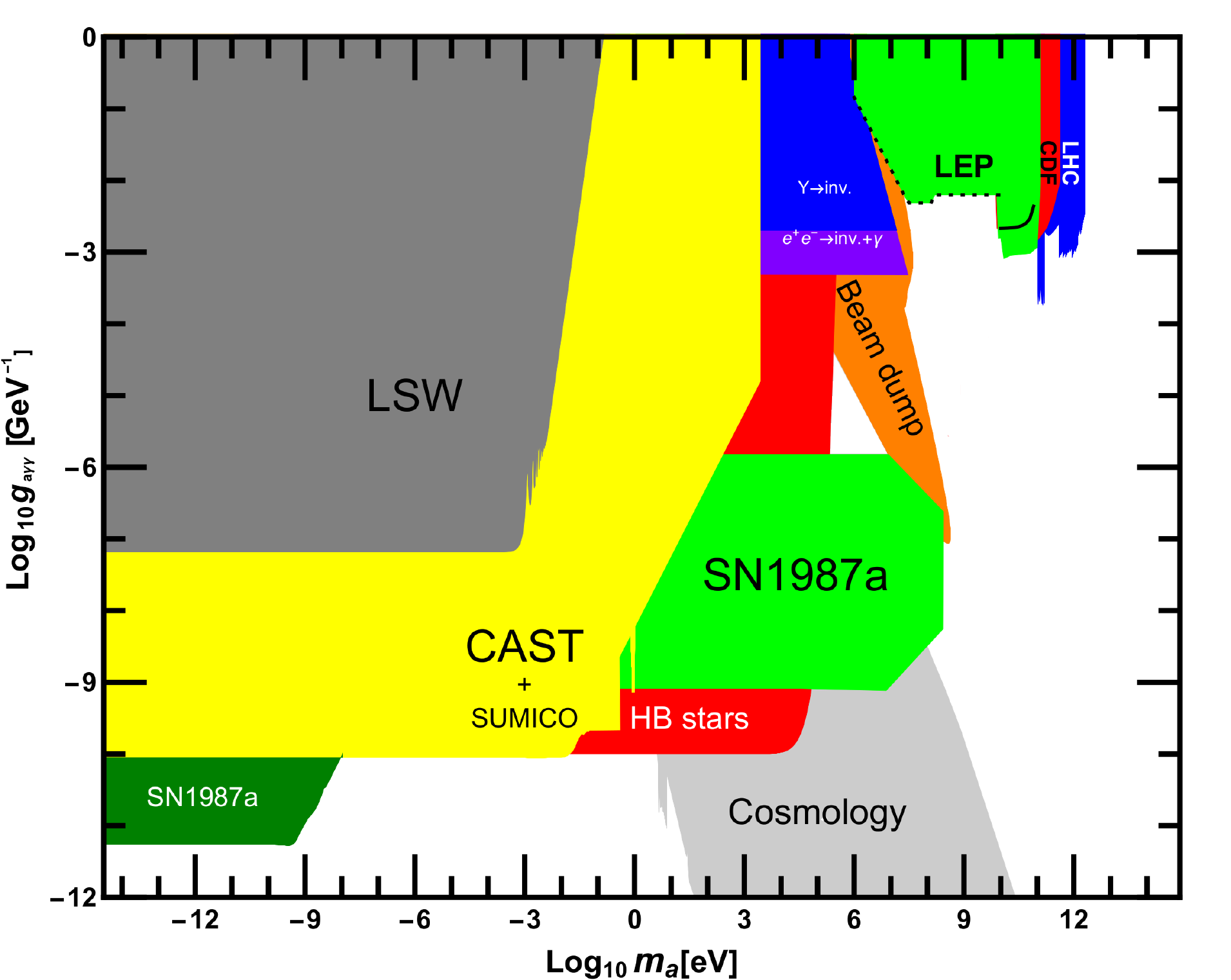}
  \caption{Limits on ALP coupling and mass parameters space compiled by Jaeckel, Jankowiak and Spannowsky~\cite{Jaeckel:2015jla} and the references therein.}
  \label{fig:ALPConstraints}
\end{figure}

When the radion's coupling to photons $g_{\sigma \gamma\gamma}$, dominates over all other couplings to standard model particles, the constraints can be recast from  ALP searches whose results are usually displayed in the mass-coupling plane~\cite{Jaeckel:2010ni,Cadamuro:2011fd,Jaeckel:2012yz,Dobrich:2015jyk,Jaeckel:2015jla}, as in Fig.~\ref{fig:ALPConstraints}. In this section we examine how these limits change when other couplings are turned on. We are primarily interested  in the region constrained by the limits from Supernova\footnote{Throughout this paper we mean by SN 1987a limit the light green region in Fig.~\ref{fig:ALPConstraints} rather than the dark green $\gamma$-burst limit.} (SN) 1987a, cosmology, Horizontal Branch stars,  and beam dump experiments \cite{Bjorken:1988as,Blumlein:1990ay,Batell:2014mga}. These bounds constrain masses in the range keV to 10 GeV, and couplings smaller than TeV$^{-1}$. These limits can be directly applied to a radion (with no other couplings) given that  scalars and pseudoscalars have very similar amplitudes\footnote{Scalars couple to ${\vec E}^2-{\vec B}^2$ while pseudoscalars couple to ${\vec E}\cdot {\vec B}$.}  for interacting with massless gauge bosons. We note that for ALPs the triangular region between beam dumps, SN 1987a, and HB stars may or may not be closed by BBN constraints, depending on further model-dependent assumptions~\cite{Millea:2015qra}. In the following, we will assume that this region is open, but will show in subsection \ref{decays}   that for masses above 1 MeV, there is a range of couplings where these additional assumptions are not needed to open this part of the window.

 In subsections \ref{SN} and  \ref{HorizontalBranch} we investigate the effects of $g_{\sigma NN}$ on the light radion window. In subsection \ref{decays} we will discuss the effects of couplings to other particles.

\subsection{SN 1987a}\label{SN}

Astrophysical objects provide a powerful natural laboratory in elementary particle physics, and stars are the best sources of weakly interacting particles such as neutrinos, gravitons, and probably radions. SN 1987a is one of the most important astrophysical sources due to its high density, high temperature, and proximity.

The light green region in Fig.~\ref{fig:ALPConstraints} shows a constraint on the coupling to photons from SN 1987a when other couplings are all neglected~\cite{Masso:1995tw}. This excluded limit covers the radion mass near MeV or less, with a coupling to photons suppressed by a scale between $10^3\TeV$ and $10^6 \TeV$.

\begin{figure}[thpb]
 \centering
  \includegraphics[height=.15\textheight]{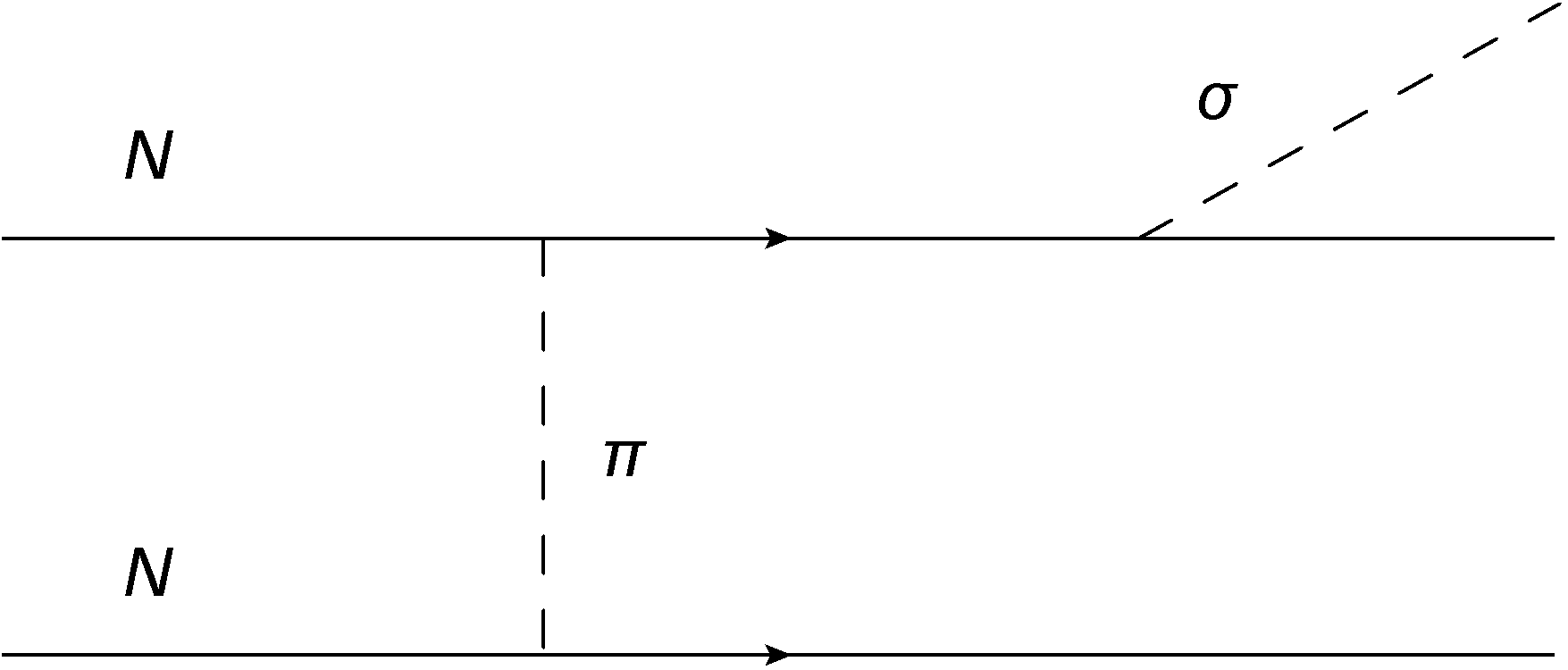}
  \caption{Nucleon-nucleon bremsstrahlung with one pion exchange.}
  \label{fig:NNbrem}
\end{figure}

Through the coupling $g_{\sigma NN}$, the radion is produced by nucleon-nucleon bremsstrahlung through an one pion exchange process. One of the eight diagrams is shown in Fig.~\ref{fig:NNbrem}. 

An approximate analytic constraint on the energy loss for SN 1987a is set by the neutrino burst duration~\cite{Burrows:1988ba} detected by IMB and Kamiokande \RN{2}. From the measured cooling rate, the energy loss rate due to beyond-the-standard-model particles should not exceed the energy loss rate through neutrinos~\cite{Raffelt:1996wa}:
\beq\label{totEbound}
\dot{E}_{\mathrm{new}} \lesssim 5\times10^{52} \erg/s.
\eeq

We will assume that the matter in the core of SN 1987a is mostly non-relativistic nucleons, i.e., $T\ll 1$ GeV. The energy loss rate per unit volume from  nucleon-nucleon bremsstrahlung ($NN\rightarrow NN\sigma$)  and the inverse mean free path of a radion in the nucleon medium due to absorption ($NN\sigma\rightarrow NN$) are given by phase space integrals of the squared amplitudes
\beq
\dot{\epsilon}=&\,\int d\Pi_1d\Pi_2d\Pi_3d\Pi_4d\Pi_\sigma (2\pi)^4 \delta^4(p_1+p_2-p_3-p_4-p_\sigma)E_\sigma      \nn\\
			&\qquad\quad \times S|\mathcal{M}_b|^2f_1f_2(1-f_3)(1-f_4), \label{epdotint}\\
\lambda^{-1} 	=&\, \frac{1}{2E_\sigma}\int d\Pi_1d\Pi_2d\Pi_3d\Pi_4 (2\pi)^4 \delta^4(p_\sigma+p_1+p_2-p_3-p_4)     \nn\\
		&\qquad \quad \times S|\mathcal{M}_a|^2f_1f_2(1-f_3)(1-f_4), \label{lambdainvint}
\eeq
where $p_1,\, p_2,\, p_3,\, p_4,$ and $p_\sigma$ are the nucleon and radion four-momenta with the subscripts 1 and 2 (3 and 4) for the incoming (outgoing) nucleons;
\beq
d\Pi_i=d^3p_i/(2\pi)^32E_i
\eeq
 is the Lorentz invariant phase-space volume element; $f_i$ are the nucleon phase-space distribution functions; and $S$ is a symmetry factor. The spin-averaged matrix element for a radion production through nucleon-nucleon bremsstrahlung in the non-relativistic limit was calculated by Ishizuka and Yoshimura~\cite{Ishizuka:1989ts}\footnote{For the free streaming limit, \cite{Ishizuka:1989ts} obtains simplified expressions for the dilaton emissivity. We note that we and the authors of \cite{Ishizuka:1989ts} agree that some of these equations were incorrect, and corrected equations are provided in Appendix~\ref{Appendix:Correction}. We thank Naruhito Ishizuka for providing the corrections.}.

The estimate for $\dot{\epsilon}$ in (\ref{epdotint}) assumes that the mean free path is much larger than the size of the region of high nuclear density. When the mean free path becomes smaller than this size, some of the radions produced in the SN will be absorbed before escaping. We can give an improved estimate of the energy loss rate due to radions that takes into account radion absorption by the following formula
\beq
\label{Eloss}
\dot{E}=\sum\limits_{i}\left[\int^{r_i}_{r_{i-1}}4\pi r^2dr~\dot{\epsilon}_i(T,\rho)\exp[-(r_i-r)/\lambda_i]\right]\prod\limits_{j>i}\exp\left[-\frac{l_j}{\lambda_j}\right]~\mathrm{erg~s^{-1}}.
\eeq
where $i$ corresponds to dividing the SN into a sequence of layers; $l_i$ is the thickness of layer $i$; $r_i$ is the distance from the center to the outmost surface of layer $i$ (i.e. $r_i\equiv \sum_{j\le i}l_j$) with the center at $r_0=0$; $\dot{\epsilon}_i$ are the energy loss rate per volume \eqref{epdotint} in layer $i$; and $\lambda_i$ is the mean free path  \eqref{lambdainvint} in layer $i$. The typical radion energy, $E_\s$, for $\lambda_i$ is chosen to be the relativistic average energy 
of a boson
\beq 
\langle E \rangle= \frac{\pi^4}{30 \zeta(3)} T
\approx 2.701\times T
\label{averageE}
\eeq in the core ($T\ge 20\MeV$) and $2.701\times20\MeV$ for outside of the core, given that the production of radions in the core dominates over the production in the outer layers.

For simplicity we assume that SN 1987a consisted of a central nucleon-rich region of four layers, with an inner core (5 km thick), an outer core (3 km), an inner mantle (10 km) and an outer mantle (10 km),  surrounded by dense gas ($\sim$1000 km) which blows off. Following the simple model in~\cite{Turner:1987by} where the temperature dependence is given as a function of the mass density: 
\beq
T(r)=(20 \MeV)(\rho(r)/10^{14} \gcmic)^{1/3}~, 
\eeq
we approximate each layer in the nucleon-rich region with a constant average nucleon mass density of $3\times10^{14}\gcmic$, $10^{14}\gcmic$, $10^{12}\gcmic$, and $10^{10} \gcmic$ with corresponding temperatures of 30 MeV, 20 MeV, 4 MeV, and 1 MeV respectively.  We numerically checked that effects from the surrounding dense gas are negligible due to its low density ($\ll 10^8 \gcmic$) and low temperature ($\ll 1 \MeV$). 

The mass density is encoded in the calculation through the chemical potential in the nucleon phase-space distribution functions. In the hot supernova core, the nucleons are partially degenerate but close to a nondegenerate state, which needs to be carefully treated. The chemical potential, $\mu$, is related to the density and temperature as \cite{Brinkmann:1988vi}
\beq
\frac{\rho}{10^{14}\gcmic}\simeq 9.0\times10^{-3}g(y)\frac{T}{\MeV}, 
\eeq
where $y\equiv(\mu-m_n)/T$, $g(y)$ in the nondegenerate ($ y \ll -1$) and degenerate limits ($ y \gg -1$) is approximated by
\beq
g(y)\simeq 
\begin{cases} \frac{\pi^{1/2}}{2} e^y , & y \ll -1, 
\\ \frac{2}{3}y^{3/2}, &  y \gg 1,
\end{cases}
\eeq
and in intermediate regime is approximated by the Taylor expansion
\beq
g(y)\simeq 0.678+0.536y+0.1685y^2+0.0175y^3-3.24\times10^{-3}y^4.
\eeq
We find $g(y)=0.3$ is a good point to divide the nondegenerate regime and the intermediate regime.

\begin{figure}[!ht]
 \centering
  \includegraphics[height=.33\textheight]{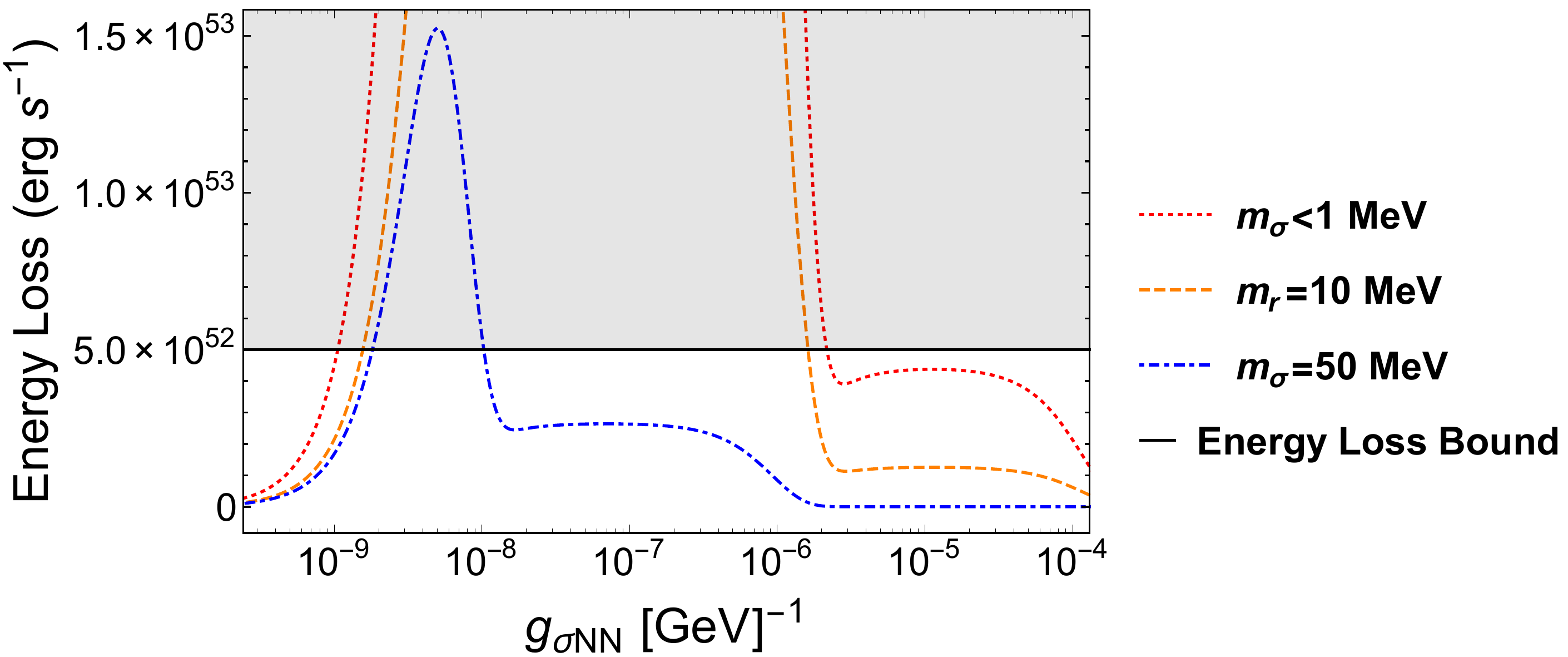}
  \caption{The energy loss rate via radions assuming they only interact with nucleons. The red-dotted line represents a radion mass $m_\s=1 \MeV$, the orange-dashed  $m_\sigma=10 \MeV$ and the blue-dot-dashed  $m_\sigma=50 \MeV$. The solid black horizontal line marks the bound ~\eqref{totEbound}.}
  \label{fig:edottotalvsgNNSN}
\end{figure}

The radion energy loss rate  \eqref{Eloss} is  plotted in Fig.~\ref{fig:edottotalvsgNNSN} along with the bound~\eqref{totEbound}.
The bump at $g_{\s NN}=1.5\times 10^{-8}\GeVinv$ is due to the discontinuity between the inner core and the outer core, and the bump at $g_{\s NN}=2\times10^{-6}\GeVinv$ is due to the discontinuity between the outer core and the inner mantle. These features would, of course, be smoothed out with a more sophisticated model of the interior. There are two regimes where the energy loss rate via radion production does not exceed the bound~\eqref{totEbound}. The first regime is where the coupling is so weak that the radions are produced too slowly to have a significant impact.  
The second regime is when the coupling is large enough that the radions cannot easily escape the SN~\cite{Burrows:1990pk}, this is the trapping regime where the radions only slowly diffuse out of the SN.  Note that when the radion mass is comparable to the typical core temperature ($\sim$ 20 MeV), the boundary of each regime is sensitive to the inner structure which is only approximately understood. 

For radions lighter than $\ll 1$ MeV, the trapping regime can be treated in another way~\cite{Raffelt:1996wa}: by calculating the luminosity of radions from a ``radionsphere" (analogous to the ``axionsphere"~\cite{Turner:1987by}) which approximates the emission by a radion blackbody. The luminosity is given in terms  of the radius $R$ of the ``radionsphere" by
\beq
L=4\pi R^2\sigma T^4(R)
\eeq
 where $\sigma$ is the Stefan-Boltzmann constant and not to be confused with the radion field. The bound on the energy loss rate (\ref{totEbound}) directly translates to a bound on the luminosity; for example for $R=$10 km, the temperature at that radius is bounded by $T(R) < 8 \MeV$. We can numerically calculate the ``radion depth" (analogous to the optical depth) from
\beq
\tau(r)=\int^{\infty}_R \lambda^{-1} dr^\prime~,
\eeq
and the radius of the ``radionsphere" is defined \cite{Raffelt:1996wa} by $\tau(R)=\frac{2}{3}$.
We checked that for the layer model described above, and for 
$g_{\sigma NN} > 10^{-6}\GeVinv$, $R$ lies in the inner mantle where the temperature is $\sim 4$ MeV, which is consistent with the more sophisticated treatment using Eq. (\ref{Eloss}). We also checked  that for the inner core \eqref{lambdainvint} yields
\beq
\lambda &\sim \frac{10^{-16}\, \textrm{km}}{(g_{\sigma NN}\,m_N)^{2}}
\eeq
for $m_\s < 50 \MeV$ in the core. This means that in the trapping regime,
the mean free path is orders of magnitude smaller than the size of the core, and radions do not alter the transfer of energy from the core.

\begin{figure}[!ht]
 \centering
  \includegraphics[height=.33\textheight]{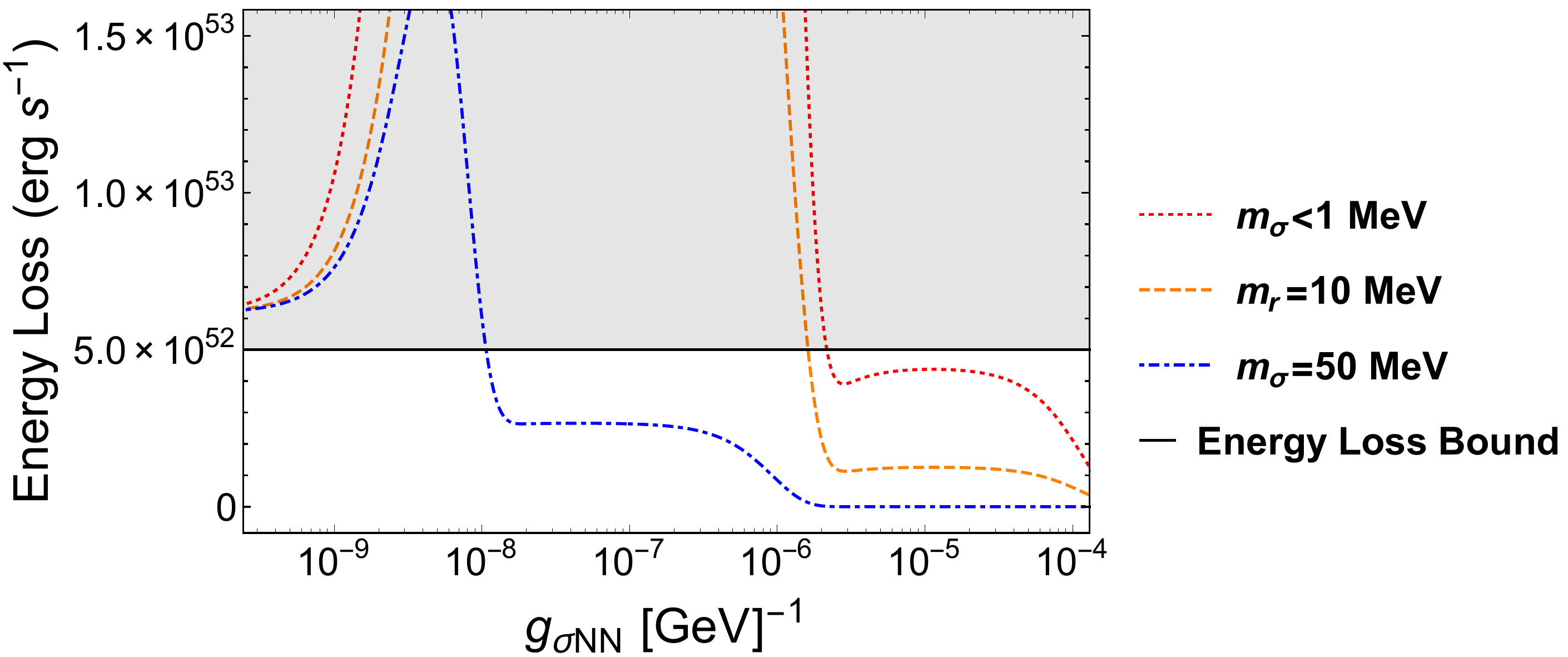}
  \caption{The energy loss rate via radions with  $g_{\s\gamma\gamma}=2\times 10^{-9}\GeV^{-1}$. See details in the text. The red-dotted line represents the radion mass $m_\s$, of 1 $\MeV$, the orange-dashed $m_\sigma=10 \MeV$, the blue-dot-dashed $m_\sigma=50 \MeV$ and the solid black horizontal line marks the bound,~\eqref{totEbound}.}
  \label{fig:edottotalwithgammavsgNNSN}
\end{figure}

\begin{figure}[!ht]
 \centering
  \includegraphics[height=.4\textheight]{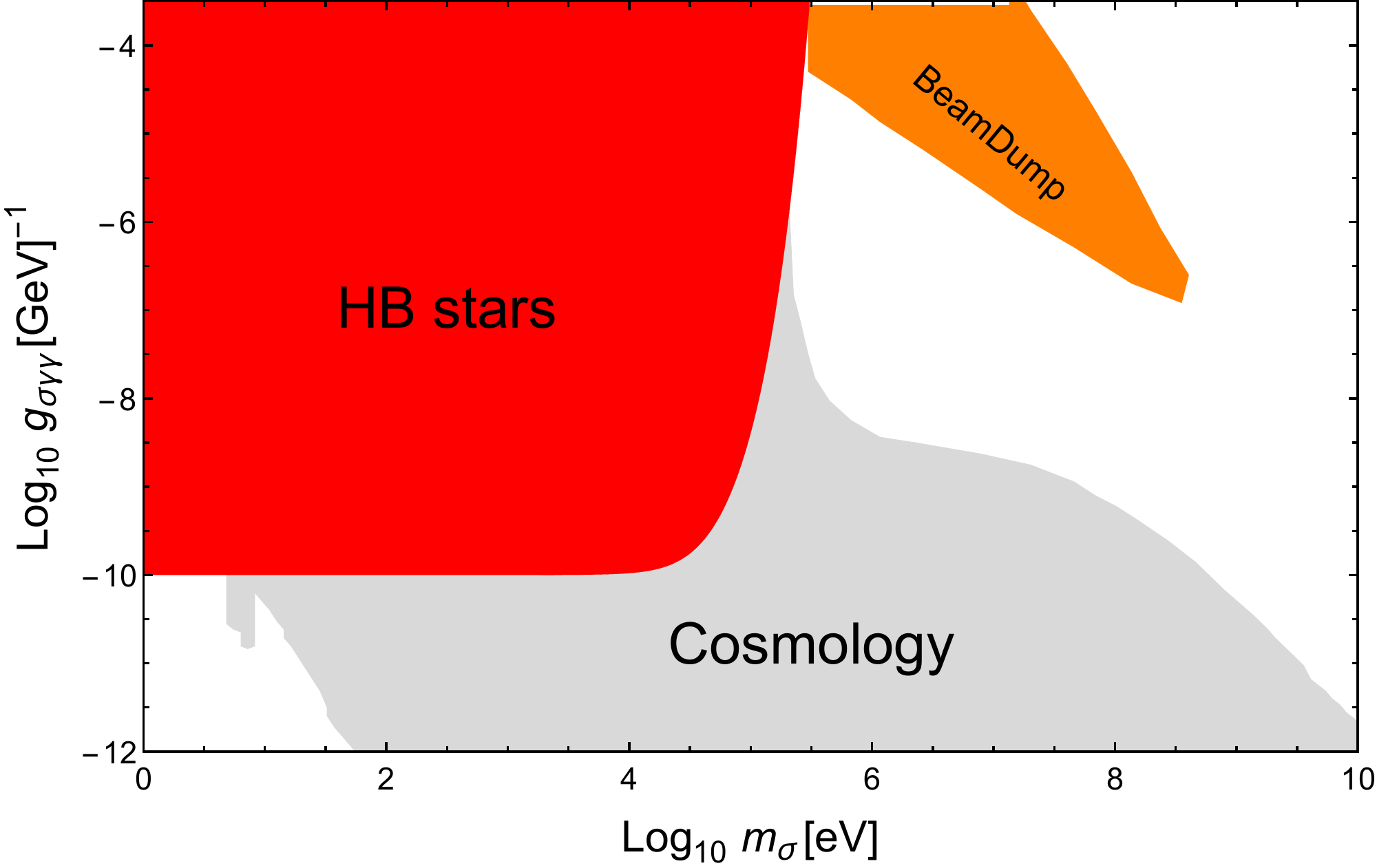}
  \caption{Limits of the coupling and the mass of the radion when $g_{\sigma NN} \gtrsim 2\times 10^{-6}\GeVinv$, modified from Fig.~\ref{fig:ALPConstraints}.}
  \label{fig:ALPlimitwoSN}
\end{figure}

\begin{figure}[!ht]
 \centering
  \includegraphics[height=.35\textheight]{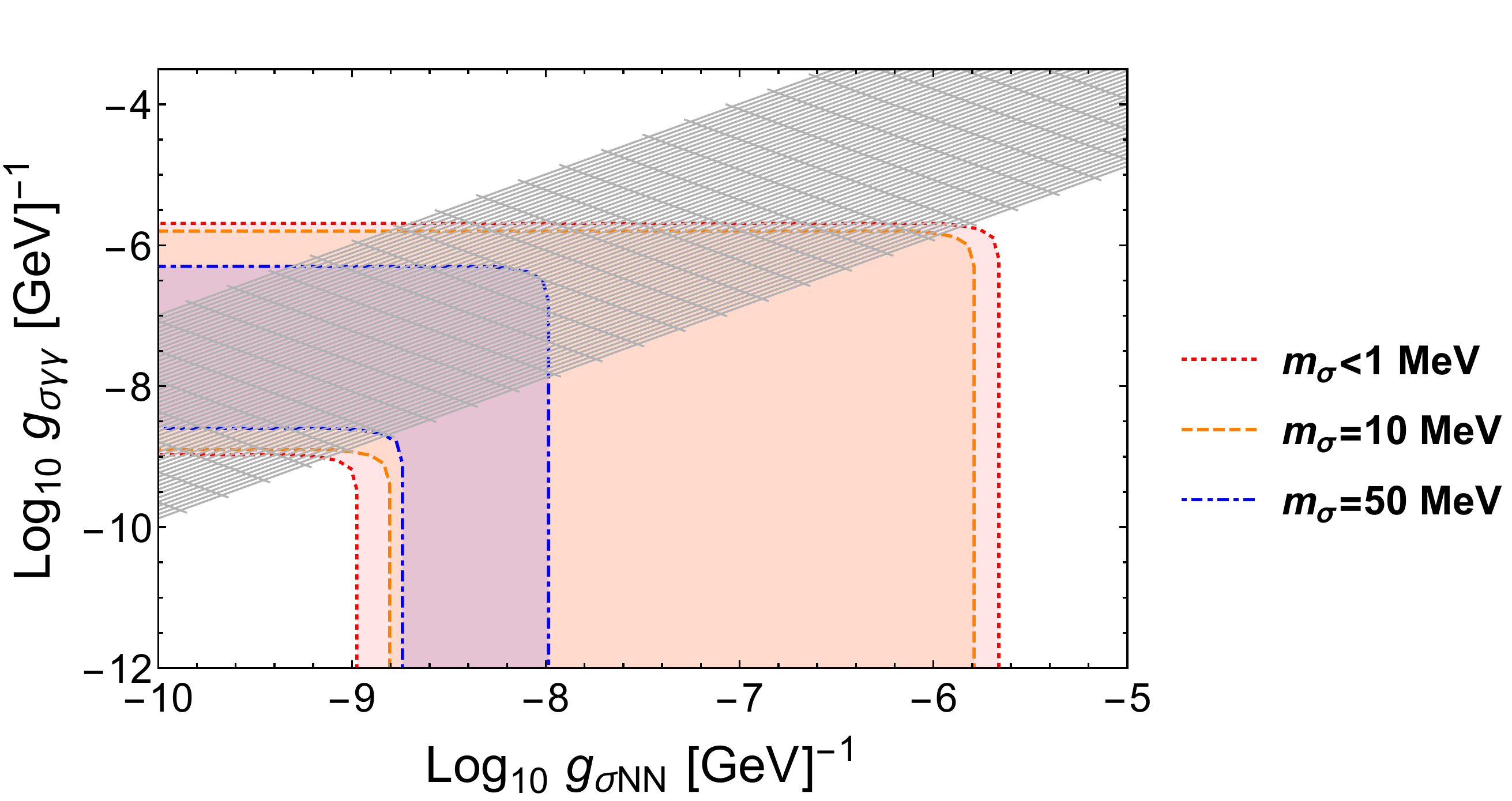}
  \caption{Exclusion region by SN 1987a in the $g_{\sigma\gamma\gamma}-g_{\sigma NN}$ space. The bound on $g_{\sigma\gamma\gamma}$ are obtained approximately from Fig.~\ref{fig:ALPConstraints}. In the case that $g_{\s\gamma\gamma}$ provides the dominant contribution to $g_{\s NN}$ the relation in Eq.~\eqref{gNggammacor}, is plotted for $0.01<|b^{EM}_{elem}+b_{IR}^{EM}|<10$ as a grey band.}
  \label{fig:ggammagN}
\end{figure}

Next we consider the case where the coupling to photons also comes into play. In Fig.~\ref{fig:edottotalwithgammavsgNNSN} the energy loss rate~\cite{Masso:1995tw} is shown, assuming an interaction strength with photons given by $g_{\s\gamma\gamma}=2\times 10^{-9} \GeV^{-1}$. We used the same layered model of densities and temperatures as above. A trapping region still remains for sufficiently large nucleon couplings.  The limit on the coupling to photons is re-plotted for the case $g_{\sigma NN} \gtrsim 2\times10^{-6}\GeVinv$ in Fig.~\ref{fig:ALPlimitwoSN}\footnote{We note that radion production through nucleon bremsstrahlung with a large nucleon coupling will modify the confidence level of the beam dump limits on the photon coupling. The specific modification is not covered in this paper. We thank Thomas Flacke for pointing this out.} where the energy loss rate goes below the bound~\eqref{totEbound}. In Fig.~\ref{fig:ggammagN} we provide the exclusion region in $g_{\sigma\gamma\gamma}-g_{\sigma NN}$ coupling space for radion masses 1 MeV, 10 MeV and 50 MeV along with a band of contours of Eq.~\eqref{gNggammacor} for $0.01<|b^{EM}_{elem}+b_{IR}^{EM}|<10$ which applies when the coupling to photons gives the dominant contribution to $g_{\s NN}$. We note that although there exist bounds from exotic meson decays~\cite{Knapen:2017xzo} on the coupling to top quarks generated radiatively from the coupling to gluons, its translation to the limit on the coupling to nucleons can be weaker depending on $b_{light}^{(3)}$ which determines a ratio of the gluon coupling~\eqref{cptogluon} to the nucleon coupling \eqref{nucleoncoupling}. We assume these exotic meson decay bounds do not affect the range we are interested in.

\subsection{The Horizontal Branch Stars}
\label{HorizontalBranch}
\begin{figure}[!ht]
 \centering
  \includegraphics[height=.35\textheight]{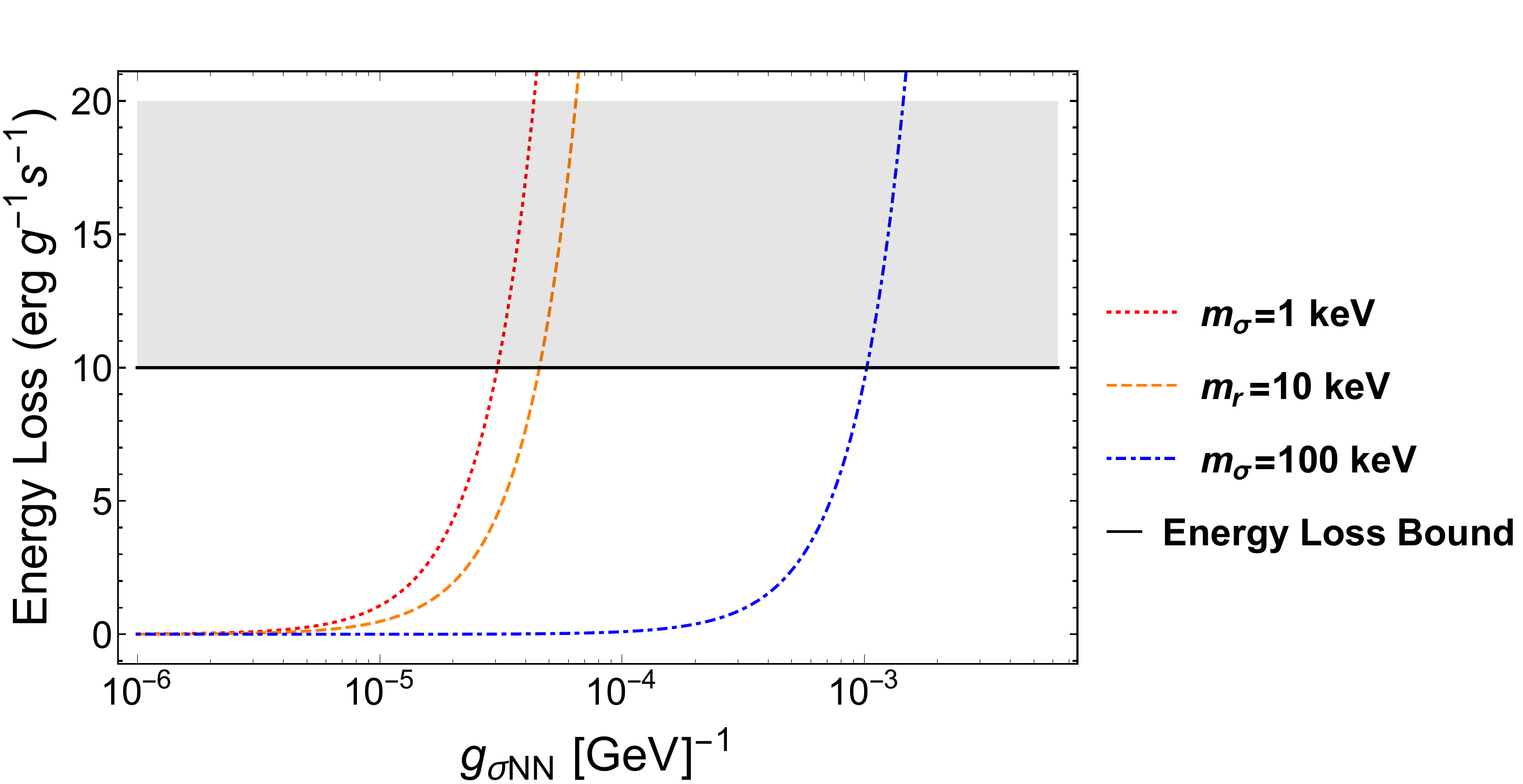}
  \caption{The energy loss rate by radions including only the interaction with nucleons. The red-dotted line represents $m_\sigma=1 \keV$, orange-dashed  $m_\sigma=10 \keV$ and blue-dot-dashed for $m_\sigma=100 \keV$. The black-solid horizontal line marks the bound ~\eqref{totEboundHB}.}
  \label{fig:ggammagNHB}
\end{figure}

Radion emission also affects the helium-burning lifetime of Horizontal Branch (HB) stars. Helium ignition can be delayed by radion cooling and this implies that the HB stars can be brighter than otherwise allowed~\cite{Raffelt:1996wa,Raffelt:2006cw}. Detailed studies~\cite{Raffelt:1996wa,Raffelt:2006cw} impose the following limit on the energy loss rate per unit mass produced by a new particle in the core,
\beq\label{totEboundHB}
\dot{\epsilon} _{\mathrm{HB}} \lesssim 10 ~\mathrm{erg\,g^{-1}\,s^{-1}}.
\eeq
A plot of the energy loss due to radion bremsstrahlung (Fig. \ref{fig:NNbrem}) with a typical core density of $\rho=10^4 \gcmic$ and a temperature $T=8.3 \keV$ corresponding to HB stars is shown in Fig.~\ref{fig:ggammagNHB} for the free streaming regime.

\begin{figure}[!ht]
 \centering
  \includegraphics[height=.35\textheight]{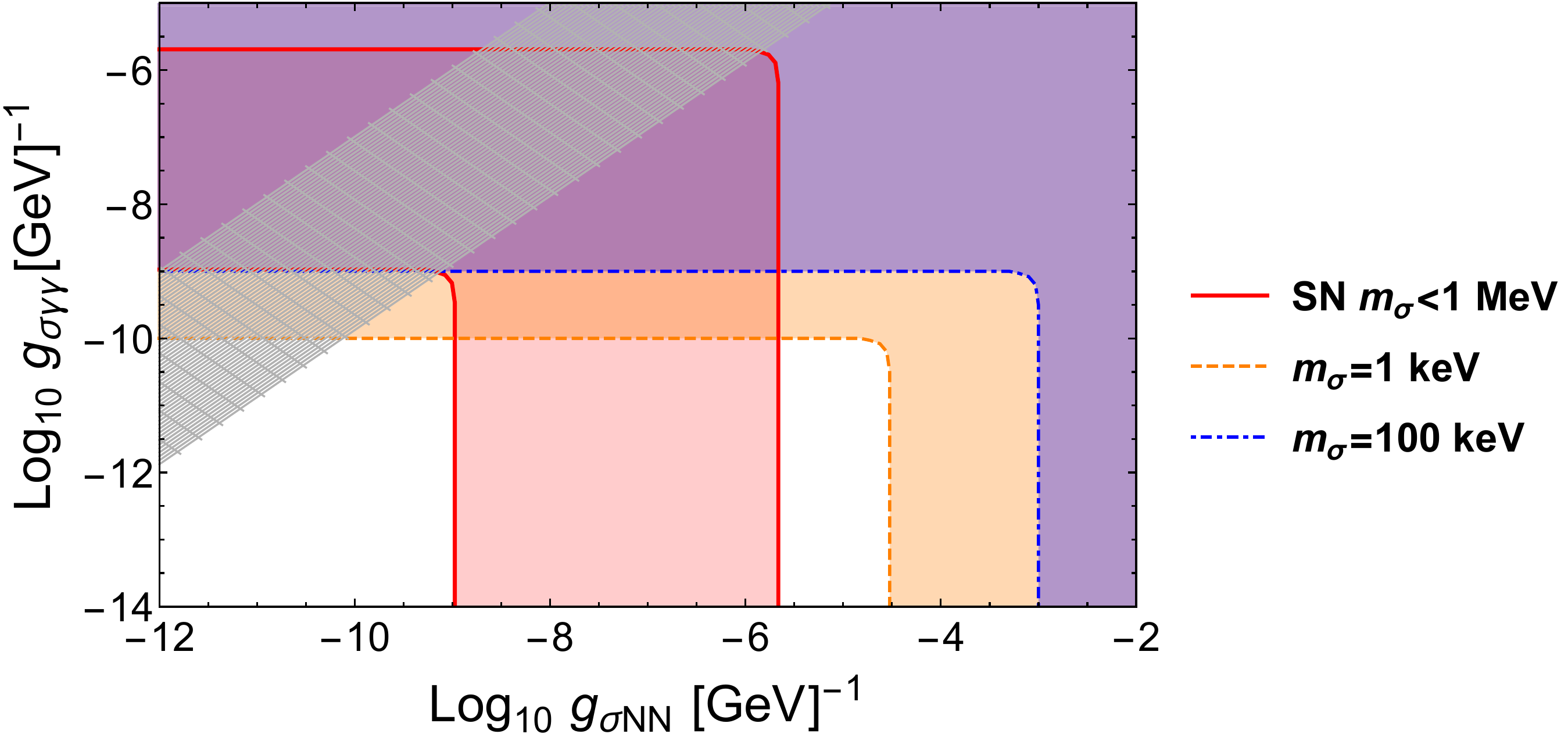}
  \caption{Exclusion region by HB in the $g_{\sigma\gamma\gamma}-g_{\sigma NN}$ space on the top of SN 1987a limit for $m_\s < 1 \MeV$. The bound on $g_{\sigma\gamma\gamma}$ are obtained approximately from Fig.~\ref{fig:ALPConstraints}. In the case $g_{\s\gamma\gamma}$ gives dominant contribution to $g_{\s NN}$ the relation, \eqref{gNggammacor} is plotted for $0.01<|b^{EM}_{elem}+b_{IR}^{EM}|<10$ as a grey band.}
  \label{fig:ggammagNHB2}
\end{figure}
Constraints on radions come from  requiring that the energy transfer by radion trapping be smaller than the radiative energy transfer~\cite{Cadamuro:2011fd,Raffelt:1988rx}. However as the dominant contribution of the energy transfer in the core of the HB stars is by convection, not by radiative transfer, this constraint should be considered conservative bound. 

Taking the typical relativistic energy (\ref{averageE}) for $m_\s \lesssim T$ or the typical non-relativistic energy, $E_\s \simeq m_\s + \frac{3}{2}T$, for $m_\s \gg T$, we numerically find that in HB stars
\beq
\lambda \sim \frac{10^{8} \, \mathrm{km} }{(g_{\sigma NN}\,m_n)^{2}},
\eeq
up to $m_\s \sim 100 \keV$, so we see that there is no possibility of a trapping regime in HB stars whose typical core radius is $10^4$ km.
If the mean free path by nucleon-nucleon bremsstrahlung could be comparable to the thickness of beam dump targets, the beam dump constraints on the two-photon coupling will be modified as radions will be trapped inside the target. Obviously, since beam dumps involve even smaller sizes, trapping will not happen and there is no change to the beam dump constraints on the two-photon coupling.

In Fig.~\ref{fig:ggammagNHB2} the exclusion region in $g_{\sigma\gamma\gamma}-g_{\sigma NN}$ plane from HB stars is shown for radion masses of 1 keV and 100 keV on the top of the SN limit for $m_\s < 1 \MeV$, again along with a band of contours of Eq.~\eqref{gNggammacor} for $0.01<|b^{EM}_{elem}+b_{IR}^{EM}|<10$ which applies when the coupling to photons gives dominant contribution to $g_{\s NN}$. There are two open windows, one at the bottom-left weak coupling limit and one at the bottom-right trapping regime where $g_{\s NN}$ bound depends on a radion mass.
%

\subsection{Limits and Radion Decays}\label{Chap:LimitsandDecays}
\label{decays}

\begin{figure}[!ht] 
  \centering
  \begin{minipage}[b]{0.49\textwidth}
    \includegraphics[width=\textwidth]{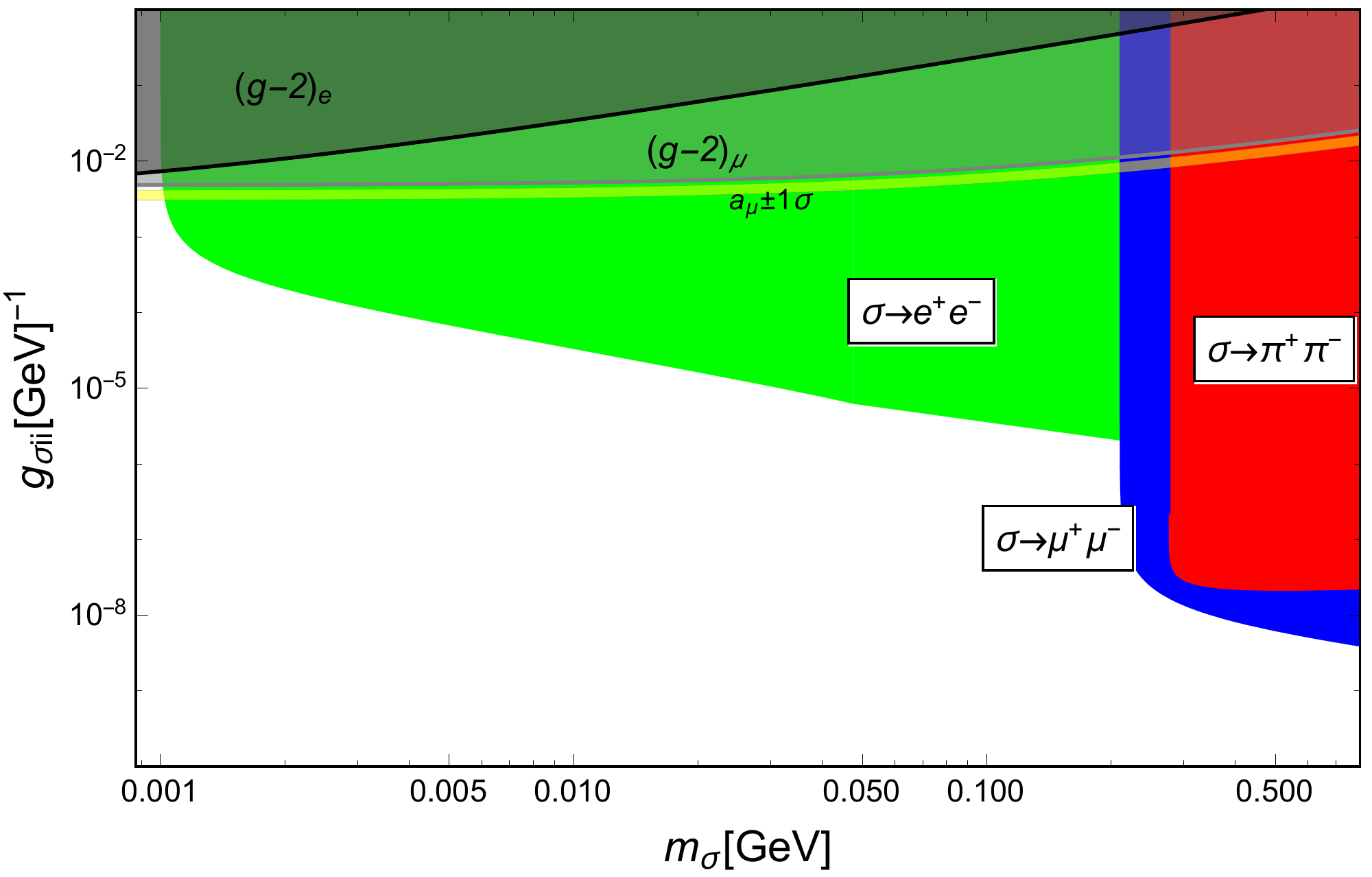}
  \end{minipage}
  \hspace{0.01cm}
  \begin{minipage}[b]{0.49\textwidth}
    \includegraphics[width=\textwidth]{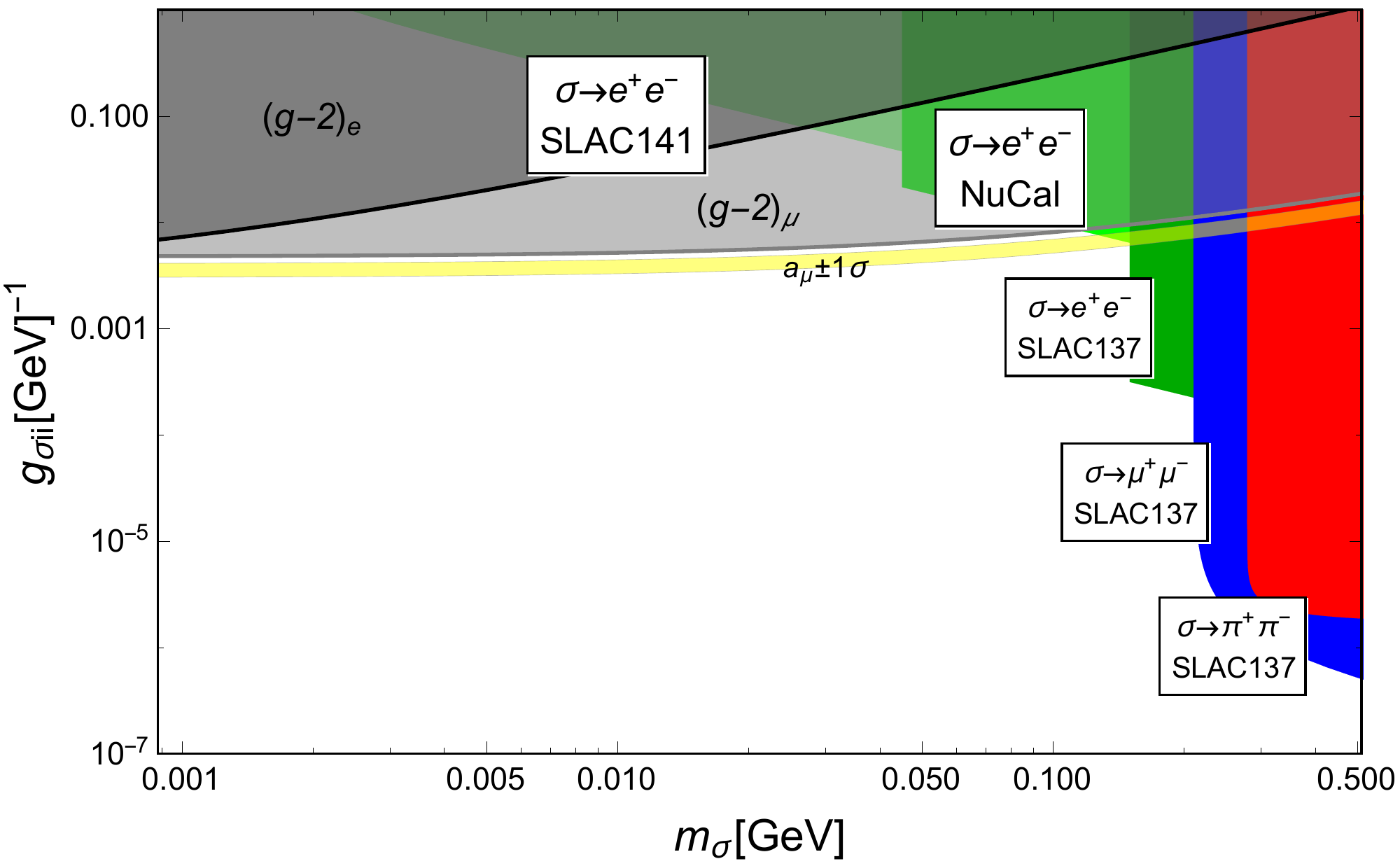}
  \end{minipage}
\begin{minipage}[b]{0.49\textwidth}
    \includegraphics[width=\textwidth]{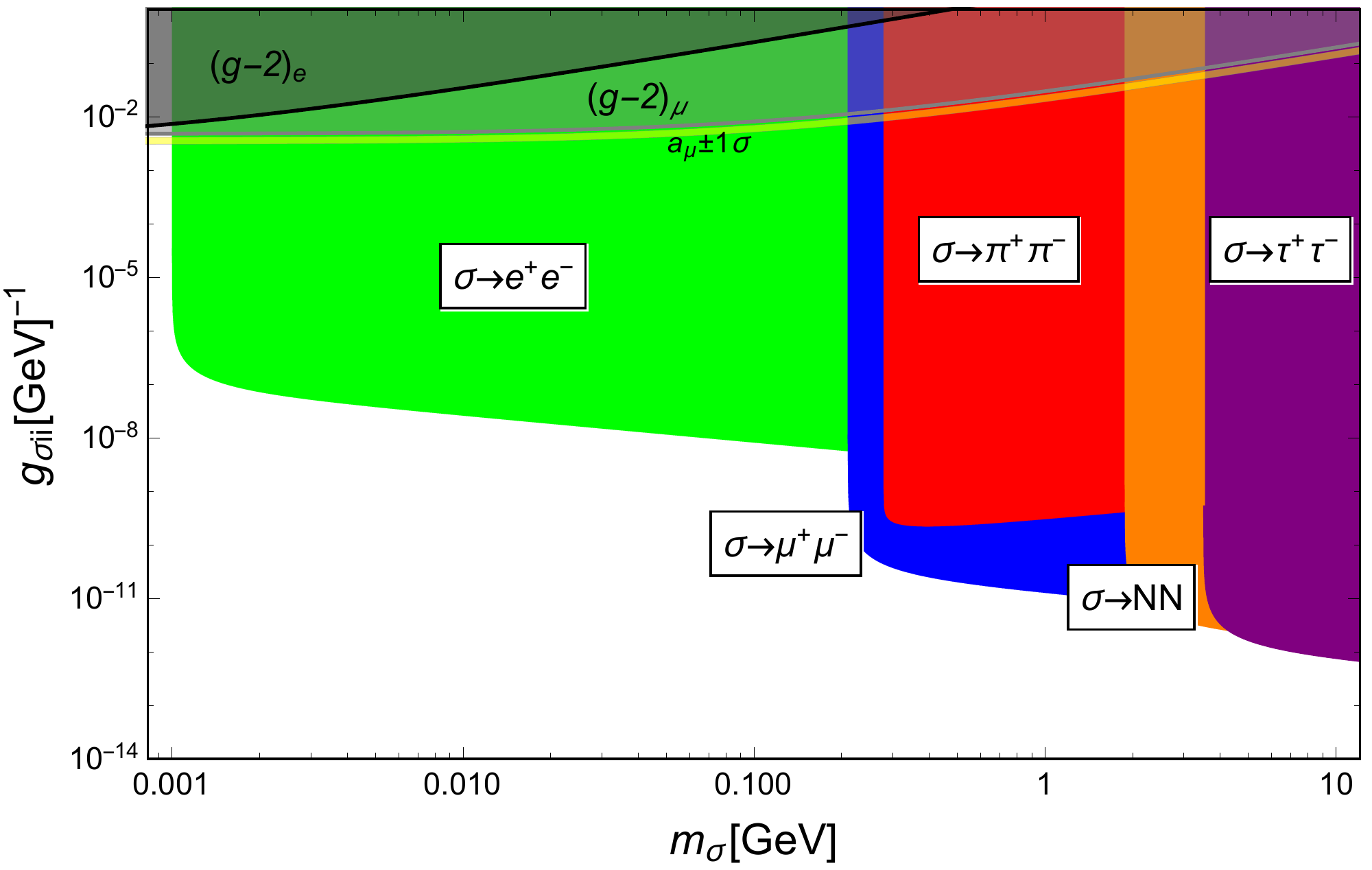}
  \end{minipage}  
      \caption{(Top Left) The region in $g_{\sigma ii} -m_{\sigma}$ parameter space where the SN 1987a bounds are affected by the radion decay to $e^{-}e^{-},\mu^{+}\mu^{-}$, and $\pi^{+}\pi^{-}$. Here we have assumed the SN has an effective radius of 15 km and a temperature of 30 MeV. (Top Right) The regions where bounds from beam dump experiments are affected by radion decay. (Bottom) The region  where the constraints from cosmological bounds are eliminated. We take the upper limit on the radion lifetime for decays to $e$'s and $\mu$'s to be 10 s, and for decays to $\pi$'s, $\tau$'s and  nucleons to be 1 s \cite{Kawasaki:2017bqm}. (All) The gray(black) regions show the excluded region from $g-2$ experiments for the muon(electron). The exclusion regions from $g_{\s ee}$ and $g_{\s \mu\mu}$ are set at 99\% confidence level, and the $\pm 1\s$ flavored band for $a_\mu$ is shown as a yellow band.}
      \label{fig:decay}
\end{figure}

Finally, we investigate the effect of radion decays on the SN 1987a,  beam dump, and cosmological bounds. If the radion  decays inside the SN or the beam dump, or before BBN, then the limits no longer apply. We consider radion decays to $e^{+}e^{-}, \mu^{+}\mu^{-}$, $\pi^{+}\pi^{-}$, $\tau^{+}\tau^{-}$, and two nucleons, and present the results in terms of the low-energy, effective couplings in (\ref{effsigmaff}) and (\ref{cptopion}). In the top left panel of Fig. \ref{fig:decay} we use Eqs.~\eqref{fermiondecay} and~\eqref{piondecay} to show  the region of the $ m_{\sigma}-g_{\sigma ii}$ parameter space where the decay length becomes smaller than the radius of the core of SN 1987a. 

 In the top right panel of Fig.~\ref{fig:decay}, we show the regions where the radion decays inside the target/absorber of various beam dump experiments. We show the experiments which give the most stringent limit for each mass range.  Note that SLAC 137  contains a hill which is an unusually long absorber \cite{Bjorken:1988as,Batell:2014mga}.

 Ref.  \cite{Kawasaki:2017bqm} provides upper bounds on lifetimes (for various decay modes) that leave BBN unaffected.  For electron and muon final states, the upper bound on the radion lifetime is 10 s, whereas the bound drops to 1 s for $\tau$'s and $\pi$'s. We show the regions where the cosmology bounds disappear in the bottom panel of Fig \ref{fig:decay}. In this case even weaker couplings can eliminate the bounds.

For radion masses larger than $2m_{e} \approx 1$ MeV, the bounds from  SN 1987a can  be modified for couplings suppressed by scales less than 
10 TeV, while the cosmology bounds can be removed with suppressions less than 100 TeV. For radion masses larger than $2m_{\mu} \approx 210$ MeV,
the beam dump bounds are relaxed with coupling suppression scales less than 1000 TeV.

\begin{figure}[!ht] 
  \centering
  \begin{minipage}[b]{0.23\textwidth}
    \includegraphics[width=\textwidth]{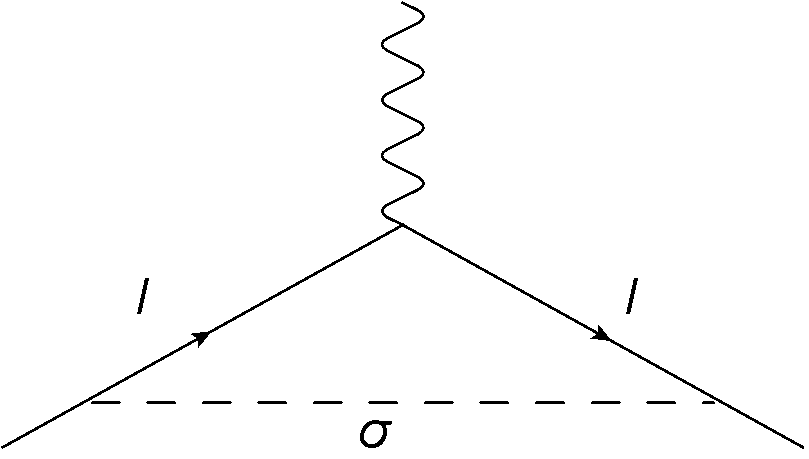}
  \end{minipage}
  \hspace{0.01cm}
  \begin{minipage}[b]{0.23\textwidth}
    \includegraphics[width=\textwidth]{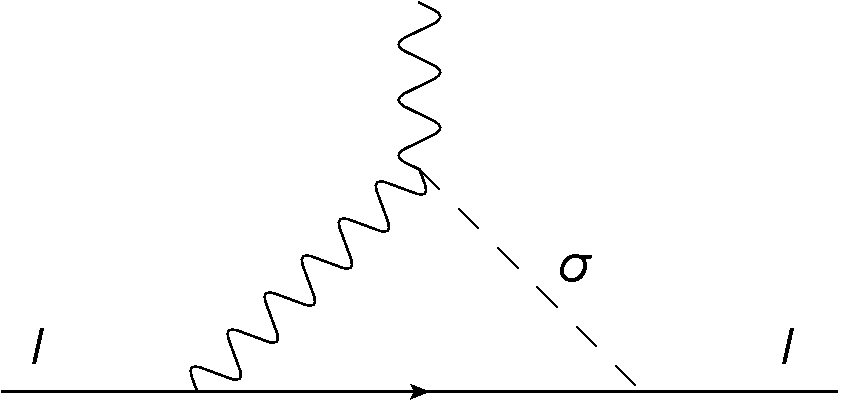}
  \end{minipage}
 \hspace{0.01cm}
\begin{minipage}[b]{0.23\textwidth}
    \includegraphics[width=\textwidth]{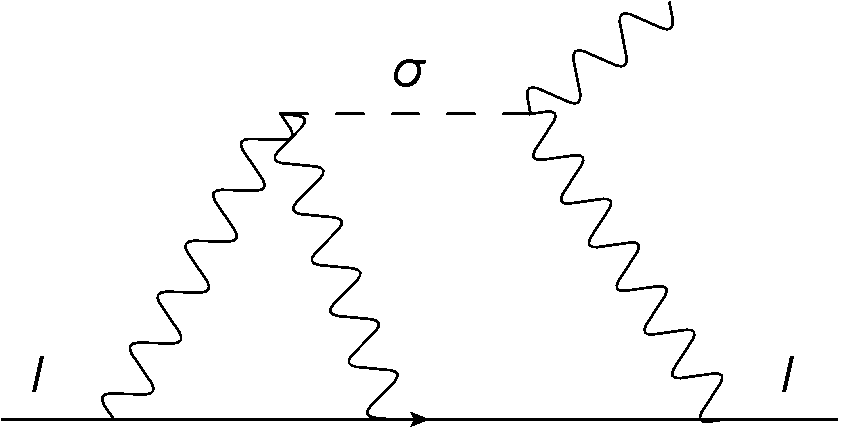}
  \end{minipage}  
\hspace{0.01cm}
  \begin{minipage}[b]{0.23\textwidth}
    \includegraphics[width=\textwidth]{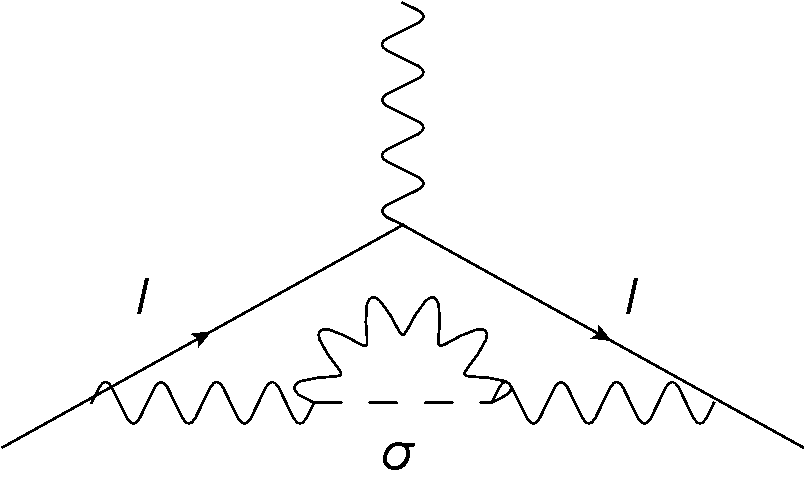}
  \end{minipage}
      \caption{LO (leftmost diagram) and NLO radion contributions to the lepton anomalous magnetic moment.}
      \label{fig:g-2}
\end{figure}

The latest experimental results of the muon anomalous magnetic moment, $a_\mu \equiv (g_\mu-2)/2$, show a 3.5$\s$ discrepancy~\cite{Chen:2015vqy,Marciano:2016yhf,deNiverville:2018hrc}:
\beq
\Delta a_\mu \equiv a_\mu^{\mathrm{exp}}-a_\mu^{\mathrm{SM}}=273(80)\times 10^{-11}.\label{muong2}
\eeq
The latest experimental results of the electron anomalous magnetic moment, $a_e$, shows a 2.4$\s$ discrepancy~\cite{Parker191}:
\beq
\Delta a_e \equiv a_e^{\mathrm{exp}}-a_e^{\mathrm{SM}}=-88(36)\times 10^{-14}.\label{electrong2}
\eeq
%
Notice that since the radion is CP-conserving, it will have no effect on the Electric Dipole Moments (EDM) of the muon and electron. Therefore EDM constraints are irrelevant in this model. The leading order (LO) and next-to-leading-order (NLO) contributions to the $\Delta a_{\mu, e}$ are shown in Fig. \ref{fig:g-2}.
The LO contribution was calculated in \cite{Chen:2015vqy}:
\beq
a_{l} = \frac{m_l^2 g_{\sigma ll}^{2}}{8 \pi^{2}} r^{-2} \int_{0}^{1} dz \frac{(1+z)(1-z)^{2}}{r^{-2}(1-z)^{2} +z} \, ,\label{g-2LO}	
\eeq
where $r \equiv m_{\sigma} / m_{l}$ for each $l=e, \mu$.  On the other hand, the NLO contributions include the Barr-Zee (BZ) contribution (second diagram in Fig. \ref{fig:g-2}), the two-loop Light-by-Light (LBL) contribution (third diagram in Fig. \ref{fig:g-2}), and the Vacuum Polarization (VP) contribution (the last diagram in Fig. \ref{fig:g-2}). All these contributions were calculated in \cite{Marciano:2016yhf} and are given by:
\beq
a_{l}^{\mathrm{BZ}} &\simeq \Big( \frac{m_{l}^2}{4 \pi^{2}} \Big) g_{\sigma \gamma\gamma} g_{\sigma ll} \ln{\frac{\Lambda}{m_{\sigma}}}\, , \label{BZ}\\
a_{l}^{\mathrm{LBL}} &\simeq \frac{3 \alpha}{\pi} \Big( \frac{m_{l} g_{\sigma \gamma \gamma}}{4 \pi} \Big)^{2} \ln^{2}{\frac{\Lambda}{m_{\sigma}}} \, ,\label{LBL}\\
a_{l}^{\mathrm{VP}} &\simeq \frac{\alpha}{\pi	} \Big( \frac{m_{l} g_{\sigma \gamma \gamma}}{12 \pi}\Big)^{2} \ln \frac{\Lambda}{m_{\sigma}}\, , \label{VP}
\eeq
where $\Lambda$ is a UV cutoff. In our calculation, we set $\Lambda = 1$ TeV. Assuming $g_{\s ii}g_{\s \gamma\gamma}$ positive, a radion gives positive contributions to lepton $g-2$. These new contributions could be a potential solution to $\Delta a_\mu$ discrepancy; on the other hand, radion contributions increase $\Delta a_e$ discrepancy\footnote{If $g_{\s ii}g_{\s \gamma\gamma}<0$, the radion contribution can be negative, hence an opposite scenario~\cite{Marciano:2016yhf}. For Fig.~\ref{fig:decay} we assume $g_{\s ii}g_{\s \gamma\gamma}>0$.}. Using these results, we can find the excluded region in the $m_{\sigma} - g_{\sigma ii}$ parameter space. We add the excluded regions corresponding to the electron and the muon to Fig. \ref{fig:decay} for the benchmark value of $g_{\sigma \gamma \gamma} = 2 \times 10^{-9} \GeV^{-1}$, with the upper $2.58\s$ deviation (99\% confidence level) from the central values of $\Delta a_{\mu,e}$,~\eqref{muong2}-\eqref{electrong2}, as the limit. Below the muon exclusion regions, the muon flavored regions are shown as a band which makes $\Delta a_\mu$ within $1\s$ deviation from the measured value.

\section{Conclusions}\label{Chap:Conclusions}

While ALPs have interesting, unconstrained regions in the mass-photon coupling plane, we have seen that radions can have either essentially the same unconstrained regions or much larger regions depending on the size of the radion coupling to other particles, especially  electrons and nucleons.
In models that realize the Contino-Pomarol-Rattazzi mechanism,
the radion mass is connected to the vacuum energy of the electroweak sector, therefore measuring the radion mass would give us indirect information about this contribution to vacuum energy. The possibility of measuring an individual sector's contribution to the total vacuum energy is  unique (at the  present time) to this class of models. 

\begin{figure}[!htb]
 \centering
      \includegraphics[width=0.7\textwidth]{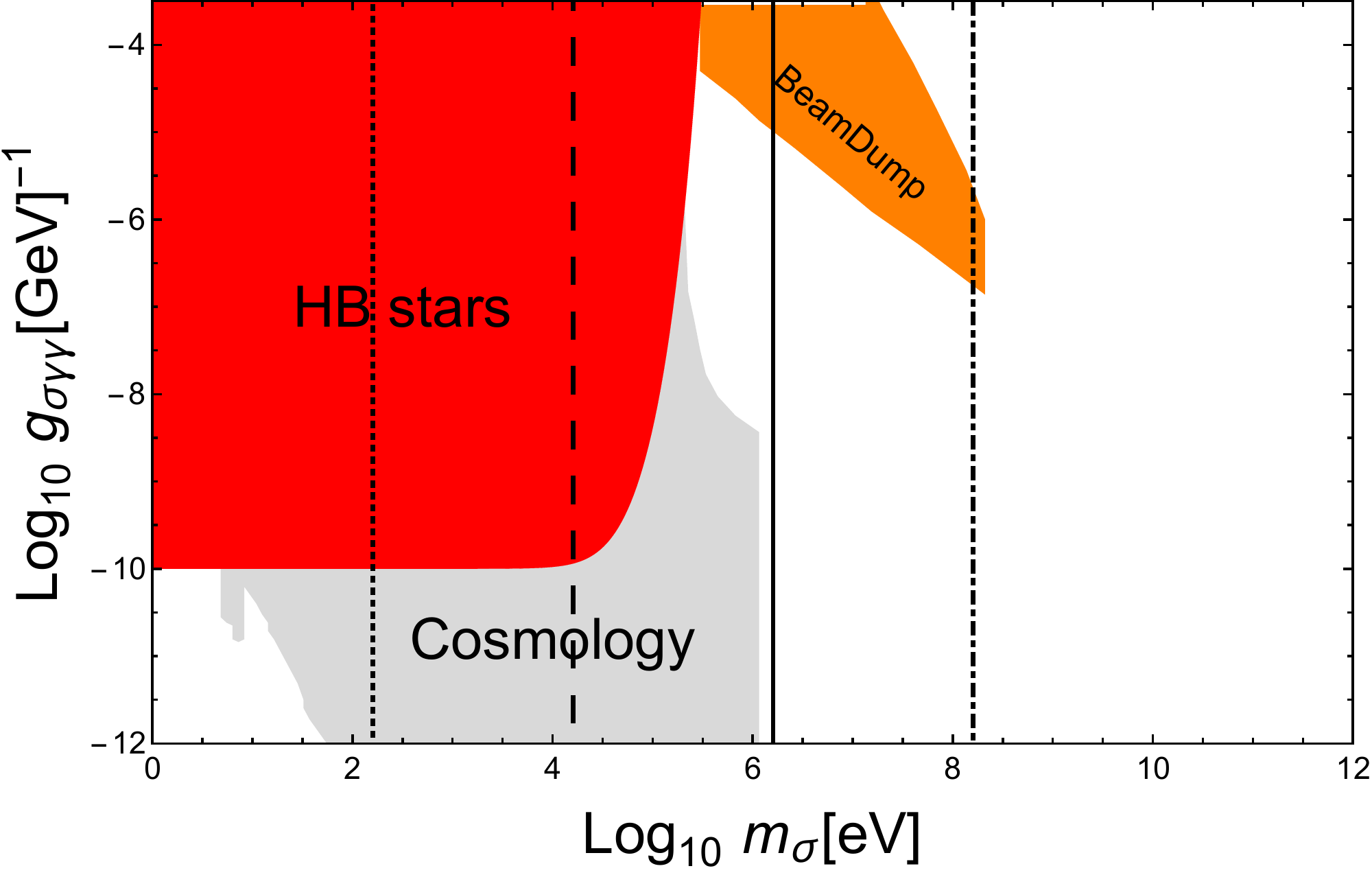}
    \caption{The open light radion window in white. The radion masses indicated by vertical lines correspond to vacuum energy contributions of $-(10\MeV)^4$ (dotted), $-(100\MeV)^4$ (dashed), $-(1\GeV)^4$ (solid) and $-(10\GeV)^4$ (dot-dashed) in a benchmark model.}
    \label{fig:vacwoSN}
\end{figure}

As an example of the kind of information one might obtain, we have overlain some radion masses that are correlated to a variety of different vacuum energies in a benchmark model\footnote{For the details of the benchmark model, see the end of Appendix~\ref{Chap:Results}.} on top of the final exclusion regions in Fig.~\ref{fig:vacwoSN}.
We have assumed that $g_{\sigma NN}$ is large enough for the radion to be in the SN trapping regime, as discussed in section \ref{SN}, so there is no constraint on the radion coupling to photons from SN 1987a. We have also assumed that $g_{\sigma ee}$ is large enough (see  Fig.~\ref{fig:decay}) so there is no constraint  from cosmology for masses above 1 MeV and that $g_{\sigma \mu\mu}$ is large enough that there is no constraint from beam dump experiments when for masses above 210 MeV. We have indicated the radion mass corresponding to electroweak vacuum energies  of $-(10 \MeV)^4$, $-(100 \MeV)^4$, $-(1 \GeV)^4$ and $-(10 \GeV)^4$. An interesting future direction would be to determine the complete range of electroweak vacuum energies that are consistent with the radion bounds in the entire class of models.


\section*{Acknowledgments}
We thank Andy Cohen, Robert Lasenby, and Martin Schmaltz for helpful discussions. We also thank Andy Albrecht and Emilja Pantic for answering our questions. This research is supported in part by DOE grant DE-SC0009999. JT thanks the Mainz Institute for Theoretical Physics (MITP) and the Aspen Center for Physics for hospitality and partial support during the completion of this work. JL thanks Naruhito Ishizuka,  Motohiko Yoshimura, Joerg Jaeckel, Michael Spannowsky, and Thomas Flacke for useful discussion and permitting to use their data and plots.

\section*{Appendix}
\appendix
\renewcommand{\thesection}{\Alph{section}}
\numberwithin{equation}{section}
\setcounter{equation}{0}
\setcounter{footnote}{0}


\section{Corrected Formulas for~\cite{Ishizuka:1989ts}}\label{Appendix:Correction}
The corresponding equation numbers in~\cite{Ishizuka:1989ts} are shown on the left in italics.
\beq
&
\textit{(18)}\quad \epsilon_D = 
\frac{ \sqrt{2} }{ 15 \cdot 32 \pi^7 } \cdot g_D^2 
\left( \frac{ f }{ m_\pi } \right)^4 
\cdot m_n^{5/2} \cdot T^{13/2} \cdot J,\\ 
& \qquad \qquad \qquad 
\cr
&
\textit{(19)}\quad J = 
\int_{0}^{\infty}
\left( \prod_{i=1}^{4} {\rm d}x_i \right) 
\theta(x_5)
\cdot x_5^2 \cdot S(x_i,y) \cdot F_D(x_i) \\
& \qquad \qquad \qquad 
\cr
&
\qquad\quad \mathrm{where\;} x_5=x_1+x_2-x_3-x_4,
\cr
\cr
&
\textit{(\textrm{B.2})}\;\; H_D(x) = \,
   86 \sqrt{x} 
+
   \frac{4}{x_5^2} \cdot 
   \Bigl[
       ( 3 \cdot c(x) - 6 \epsilon^2 ) \cdot x^{1/2}   
     + (         d(x) - 4 \epsilon   ) \cdot x^{3/2} 
     - \frac{3}{5}                     \cdot x^{5/3} 
   \Bigr]
\cr
& \qquad\qquad\quad
+ 
   \sqrt{\epsilon}\cdot {\rm arctan} ( \sqrt{x/\epsilon} ) \cdot
   \Bigl[
       -142 + \frac{4 \epsilon}{x_5^2} 
              \cdot \bigl\{
                       5 \epsilon + 7 \cdot d(x) - 5 \cdot c(x) / \epsilon
                    \bigr\}
   \Bigr]
\cr
& \qquad\qquad\quad
+ 
  \epsilon \frac{\sqrt{x}}{x+\epsilon} \cdot
  \Bigl[
     30 - 4 \epsilon/x_5^2
              \cdot \bigl\{
                       \epsilon + d(x) - c(x) / \epsilon
                    \bigr\}
  \Bigr]
\cr
& \qquad\qquad\quad
+ I(x)
     \cdot \bigl\{
             13 \epsilon^2 + 4 \epsilon^4 / x_5^2 - 31/4 \cdot x_5^2 
                    \bigr\},\\
\cr
&
\qquad\quad \mathrm{where\;} c(x) = (x_4-x_2)(x_3-x_1),\;  d(x) = x_1 + x_2 + x_3 + x_4, \mathrm{\;and\;}
\cr
\cr
&
I(x) 
= \frac{\theta(Z)}{\sqrt{Z}} \cdots
\cr
& \qquad \quad
( \mbox{same function as $\{ \frac{\theta (Z)}{\sqrt{Z}} \cdot \ln \cdots + \frac{\theta(-Z)}{\sqrt{-Z}} \cdot {\rm arcsin} \cdots \}$ }
\cr
& \qquad\qquad \qquad 
   \mbox{ appeared in 2nd and 3rd line in \textit{(A.17)} of~\cite{Ishizuka:1989ts}}).
\eeq
The definitions not specified are the same as in~\cite{Ishizuka:1989ts}.

\section{Bulk Gauge Bosons}\label{massivegbmass}

The action for a bulk gauge field is
\beq \label{MassiveGB}
S_{gauge} = \int d^{4}x dy \sqrt{g} g^{MP}g^{NQ}\Big[-\frac{1}{4g_{5}^{2}} W_{MN}^{a}W_{PQ}^{a} \nn \\ 
+\frac{v^{2}}{8}\sqrt{g_1} \delta(y-y_{1}) g^{MP} \Big(W_{M}^{a}W_{P}^{a}   \Big)    \Big]\, ,
\eeq
where $v$ is localized Higgs VEV on the IR brane. Taking the KK decomposition
\beq \label{KKDecomposition}
W_{\mu}^{a}(x,y) = W^{(n)}_{\mu}(x)f^{(n)}(y)\, ,
\eeq
where $f^{(n)}(y)$ satisfies the normalization condition
\beq \label{KKDecomposition2}
\frac{1}{g_5^2}\int dy f^{(n)}(y)f^{(m)}(y)=\delta^{\mn}\frac{1}{g_4^2},
\eeq
the equation of motion is
\beq \label{WEOM}
\Big(e^{-2A}\partial_{y}^{2} - 2A^\prime e^{-2A} \partial_{y} + m_{n}^{2} -\frac{v^2g_5^2}{4}e^{-2A}\delta(y-y_1) \Big) W_{\mu}(y) = 0.
\eeq
with  a solution for the background metric $A(y)$ as in~\eqref{bgsolmetric}. Fixing the mass of the lightest mode, which corresponds to the $W$ boson, determines the Higgs VEV $v$. The KK masses are obtained as the eigenvalues $m_n$, of the KK towers  in  \eqref{WEOM}.

The 4D effective gauge coupling is matched to the 5D gauge coupling by integrating out the extra dimension at tree-level:
\beq
\begin{split}
\mathcal{L}&\supset \int dy \sqrt{g}(-\frac{1}{4g_5^2}F_{MN}F^{MN})\\
		&=2\int^{y_1}_{y_0} dy (e^{-4A})(-\frac{1}{4g_5^2}F_{\mu\nu}F^{\mu\nu}e^{4A}) \\
		&\equiv-\frac{1}{4g_4^2}F_{\mu\nu}F^{\mu\nu}\, ,
\end{split}
\eeq
where for the massless zero mode $\partial_y A_{\mu}(x,y)=0$ and $A_5=0$ are the gauge fixing conditions. The fields is the third line are contracted with the Minkowski metric and the orbifolding factor of 2 is explicitly included on the second line. As the $F^2$ term in the second line doesn't depend on $y$, we then have
\beq
\int dy (-\frac{2}{4g_5^2})&=-\frac{1}{4g_4^2}\, , \nn\\
\frac{2L}{g_5^2}&=\frac{1}{g_4^2}\, , \label{running}
\eeq
where $L\equiv y_1-y_0$ is the size of the extra dimension. The energy scale is given by $\mu= ke^{-A(y)}$, where $k$ represents the curvature scale near the UV brane. In AdS we have $A(y)=k y$, and
\beq
L=y_1-y_0=\frac{1}{k}\log(\frac{\Lambda}{f})\, .
\eeq

The CFT contribution to the $\beta$-function is parameterized in terms of bulk parameters as
\beq
\beta(g)=&\frac{\partial g}{\partial \log\mu}=\frac{1}{-2g^{-3}}\frac{\partial }{\partial \log\mu}\left(\frac{\frac{2}{k}\log(\frac{\Lambda}{\mu})}{g_5^2}\right)=\frac{g^3}{kg_5^2}\equiv -\frac{b_{CFT}g^3}{2(8\pi^2)}~.
\eeq


\section{Model Results}\label{Chap:Results}

Here we will show some numerical values calculated with different benchmark parameter sets. 
 The minimum of the effective potential determines the hierarchy, so that $k \Braket{\chi} \sim$ TeV.  For numerical simulations we use a parameter $\alpha$ to specify the hierarchy between the UV and IR:
 \beq
k \Braket{\chi} = \alpha\TeV\, .
\eeq
This sets the scale factor $k$ and thus determines the masses in the model. We give the masses in units of $k$ (i.e.  if the dimension is [mass]$^2$ then it is given in units of $k^2$, etc.) for all other parameters unless otherwise specified. Throughout this paper we fix $\kappa=0.5$, $\lambda_{0,1}=10^{30}$ and $\mu_0=1$ (i.e. $y_0=0$). 

\begin{figure}[htb]
 \centering
  \includegraphics[height=.26\textheight]{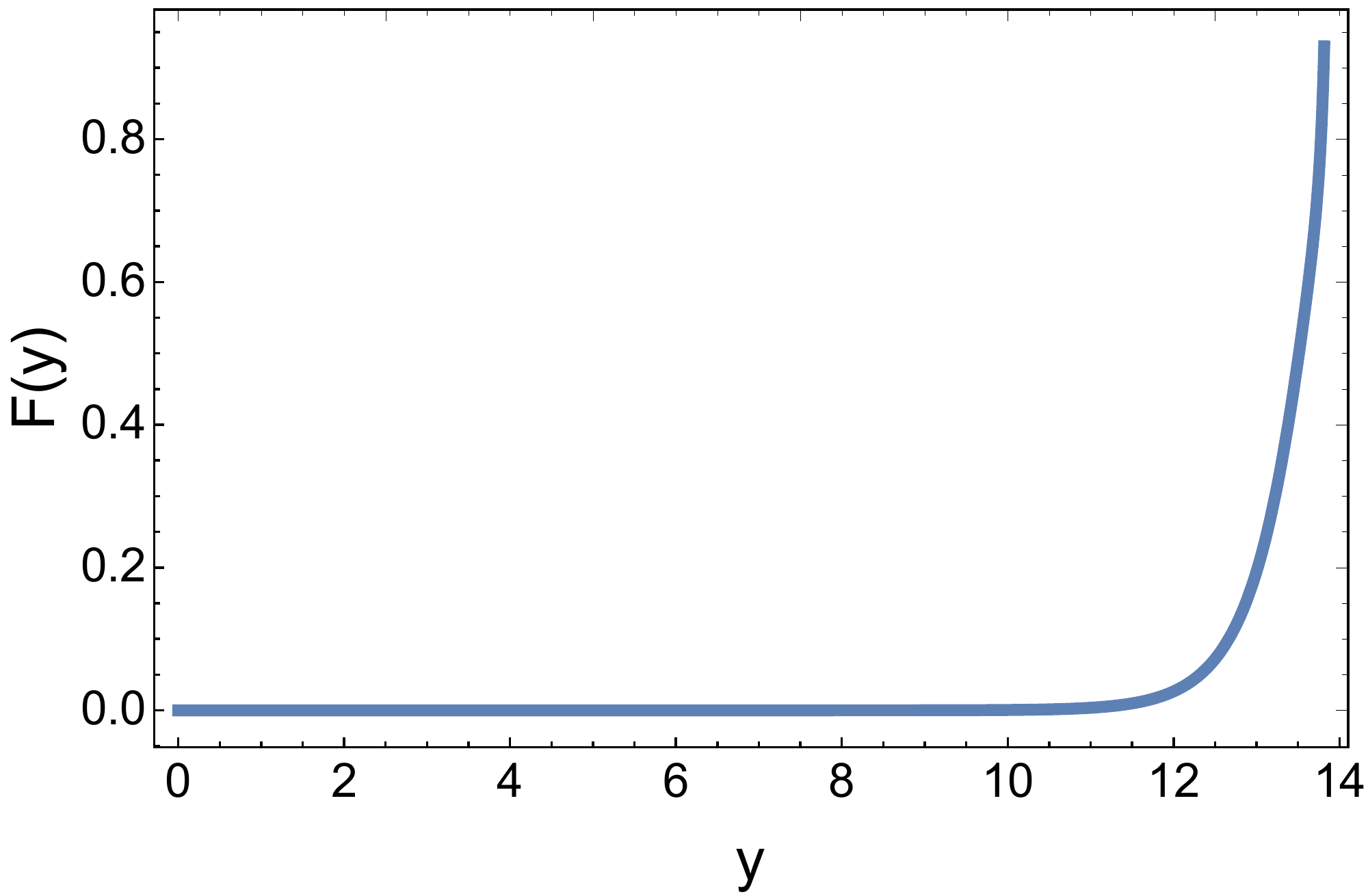}
  \caption{The bulk profile of the (unnormalized) fluctuation of the metric $F(y)$ in \eqref{perturbedmetric}. $y_1=13.8,\; v_1=5, \; T_1=-60, \;\eps=10^{-13}$.}
  \label{fig:FVSy}
\end{figure}

In Table \ref{tab:radionmass}, we display the contribution to the vacuum energy ($V^{IR}_{min}$) with the mass of the radion for each $\eps$ for two different benchmark parameter sets. We can see that $V^{IR}_{min}$ is proportional to $\eps$ and $m_{radion}$ is proportional to $\eps^{1/2}$ as expected. Numerically we also check that the vacuum and the mass are not sensitive to the bulk parameter $y_1$\footnote{We note that $y_1$ is controlled by $v_0$ when the other parameters are fixed, and due to the large value of $\lambda$, we are making $y_1$ shifts through changes to a high number of digits of $v_0$.\label{fn:y1}} whereas they are sensitive to the IR brane parameters $T_1$ and $v_1$. This can be understood by noting that the metric fluctuation peaks near the IR brane (Fig.~\ref{fig:FVSy}).

\begin{table}[!ht]
  \centering
  \begin{tabular}{c|c|c}
    \multicolumn{3}{c}{a) $y_1=13.8,\; v_1=3, \;T_1=-60$} \\
    \hline \hline
	$\eps$   	& $V^{IR}_{min}$$\;(\GeV)^4$	 & $m_{\sigma}$ (MeV) \\ \hline 
    $10^{-17}$ & -0.0000130 &0.00284\\
    $10^{-15}$ & -0.00130 &0.0284\\
    $10^{-13}$ & -0.130 &0.284\\
    $10^{-11}$ &-13 &2.84\\
  \end{tabular} \hspace{0.5cm}
  \begin{tabular}{c|c|c}
    \multicolumn{3}{c}{b) $y_1=13.8,\; v_1=5,\; T_1=-40$} \\
    \hline \hline
	$\eps$   	& $V^{IR}_{min}$$\;(\GeV)^4$	 & $m_{\sigma} $ (MeV) 	\\ \hline
    $10^{-17}$ & -0.000107 & 0.00828	\\
    $10^{-15}$ & -0.0107 &0.0828	\\
    $10^{-13}$ & -1.07 &0.828		\\
    $10^{-11}$ & -107 &8.28		\\
  \end{tabular} 
  \caption{The contribution to the vacuum energy and the radion mass for each benchmark parameter set, with $\alpha$=1. Different values can be obtained by rescaling the values by $\alpha$. The tuning between UV and IR value of the field  $\phi$ given by $v_0/v_1$, is equal to 0.095 for a) and 0.62 for b) where the changes of $v_0$ for different $\epsilon$ values take place at a high number of digits.}
  \label{tab:radionmass}
\end{table}

In Table \ref{tab:Wprime}, we provide the mass of the lightest KK mode of the $W$ boson, $m_{W^\prime }$, and $f$ defined in \eqref{cptogluon} with two different parameter sets (see Appendix~\ref{massivegbmass} for a discussion of mass of the KK gauge boson).  We also check that the mass of the lightest KK $W$ boson and  $f$ are not sensitive to $\eps$ when $\eps$ is smaller than $10^{-1}$. This can be understood by the fact that the mass of the KK $W$ boson mainly comes from the bulk gradient contribution, while the Higgs mechanism on the IR brane contributes very little.

\begin{table}[!ht]
  \centering
  \begin{tabular}{c|c|c}
    \multicolumn{3}{c}{a) $y_1=13.8,\; v_1=5 ,\; \eps=10^{-13}$} \\
    \hline \hline
	$T_1$    & $m_{W^\prime }$ (TeV) & $f$ (TeV)\\ \hline
    $-60$ 	&2.5537	& 140.37\\
    $-50$  	&2.5548  & 139.44\\
    $-40$  	&2.5561  & 138.24 \\
 
  \end{tabular} \hspace{1cm}
  \begin{tabular}{c|c|c}
    \multicolumn{3}{c}{b) $y_1=16.1,\; v_1=5,\; \eps=10^{-13}$} \\
    \hline \hline
	$T_1$    & $m_{W^\prime }$ (TeV)  & $f$ (TeV)\\ \hline
    $-60$ 	&2.5428	& 163.76\\
    $-50$  	&2.5438	& 162.67\\
    $-40$  	&2.5450	& 161.27\\
  \end{tabular} 
  \caption{The mass of the lightest KK mode of the $W$ boson and the radion VEV for two benchmark parameter sets, with $\alpha$=1. The  tuning $v_0/v_1$, for  $T_1$ = -60, -50 and -40 is 0.46, 0.53 and 0.62 respectively for both a) and b).}
  \label{tab:Wprime}
\end{table}

\begin{figure}[!ht]
    \subfloat[$x$ vs $\alpha$ \label{subfig-1:dummy}]{%
   \includegraphics[height=.23\textheight]{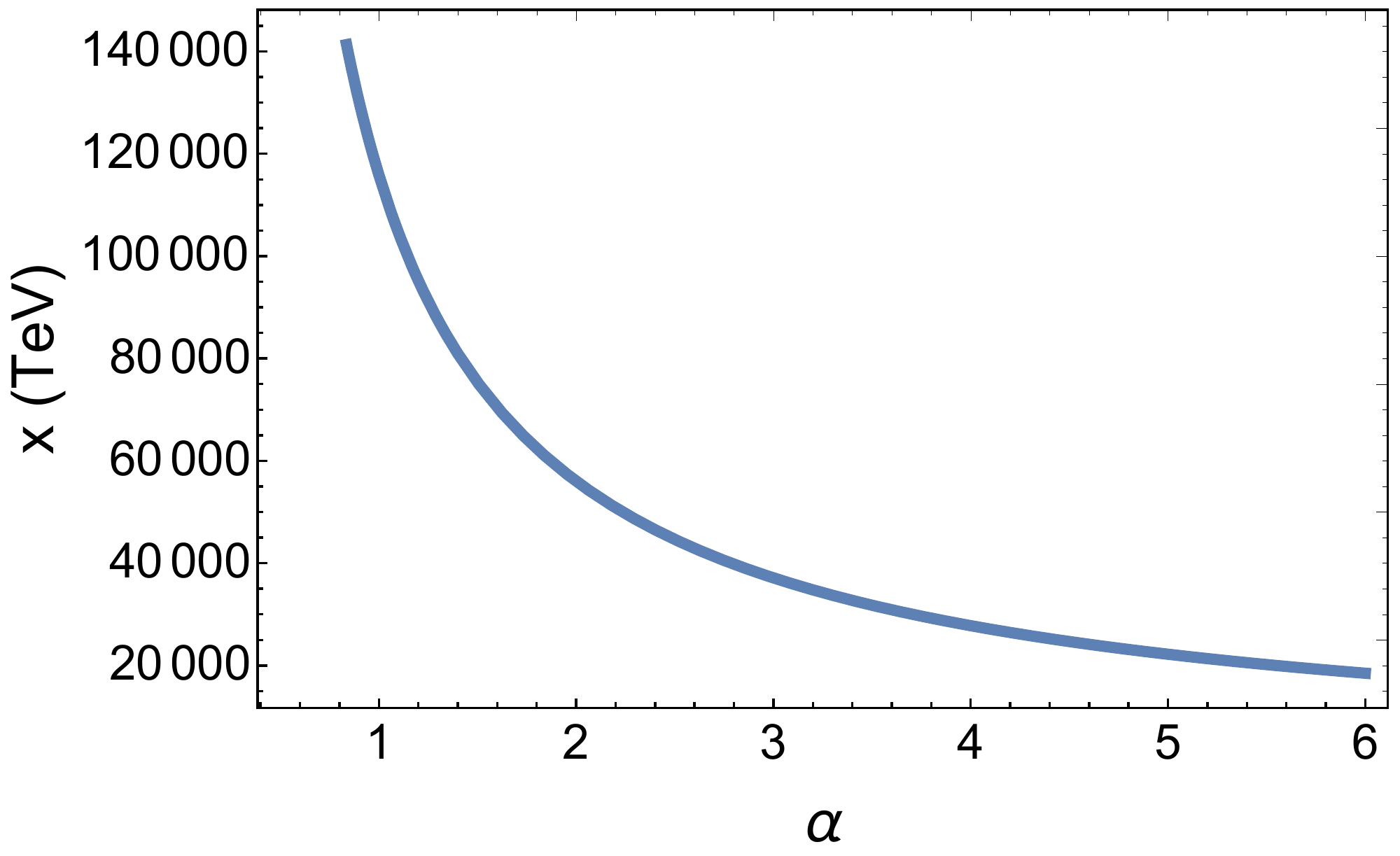}

    }
    \hspace{0.1cm}
    \subfloat[$W^\prime$  mass vs $\alpha$\label{subfig-2:dummy}]{%
  \includegraphics[height=.23\textheight]{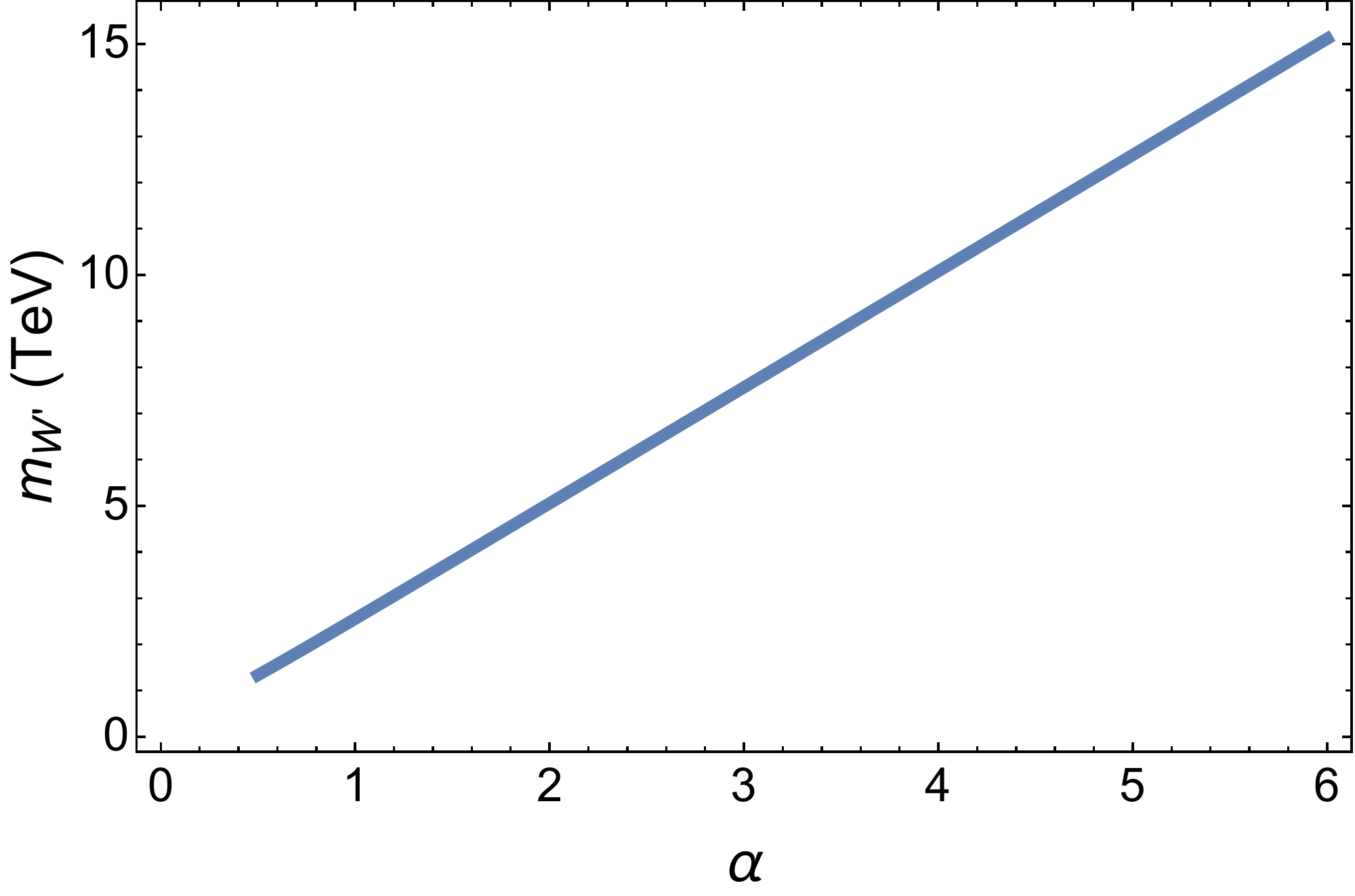}
    }
    \caption{Plots of $x=v^2g_5^2/4$ (left) and $m_{W^\prime }$  (right) versus $\alpha$. $y_1=13.8,\; v_1=5, \; T_1=-50, \;\eps=10^{-13}$. }
    \label{fig:vsdimscale}
\end{figure}

Special attention has to be paid to the scale factor $\alpha$. If we change this scale factor, then all the masses in Table \ref{tab:radionmass} are simply multiplied by $\alpha$. The results in Table \ref{tab:Wprime}, however, are not obtained by simply multiplied by $\alpha$ because we need to set the $W$ mass to 80 GeV. However, due to the flatness of the bulk wavefunction, its mass comes mainly from the Higgs VEV while the mass of the KK mode mainly comes from the bulk gradient. The Higgs VEV has to be adjusted depending on the value of $\alpha$. In Fig.~\ref{fig:vsdimscale} we show how $x\equiv v^2g_5^2/4$ and $m_{W^\prime }$ vary as $\alpha$ varies, where $v$ is the Higgs VEV that arises on IR brane. We can see that the proportionality between $m_{W^\prime }$ and $\alpha$ is preserved and $x$ gets smaller as $\alpha$ increases to preserve the zero mode mass. This scale factor parameter allows the model to easily escape a lower bound on the mass of the KK modes coming from the experiments as long as the bound is not much larger than $\order{10}$ TeV if we want to keep $k \Braket{\chi} \sim O(10)\TeV$

To make Fig. \ref{fig:vacwoSN} we used $T_1=-40$, $v_1=5$, $y_1=13.8$, $\alpha=0.5$
with different values of $\epsilon$ to achieve different vacuum energies.


\end{document}